\documentclass[pre,aps,twocolumn,superscriptaddress,longbibliography,notitlepage]{revtex4-2}

\usepackage[colorlinks,linkcolor=darkblue,citecolor=darkblue,urlcolor=darkblue]{hyperref}
\usepackage{microtype}
\usepackage{mathptmx}
\usepackage{amsfonts}
\usepackage{color}
\usepackage{graphicx}
\usepackage{float}
\usepackage{bm}
\usepackage{bbm}
\usepackage{physics}
\usepackage{mathtools}
\usepackage{appendix}
\usepackage{booktabs}
\usepackage[utf8]{inputenc}
\usepackage{multirow}

\newcommand{\beq}{\begin{equation}}
\newcommand{\eeq}{\end{equation}}

\definecolor{deepblue}{rgb}{0,0,0.5}
\definecolor{apsblue}{rgb}{0.18,0.19,0.57}
\hypersetup{
colorlinks,%
citecolor=apsblue,%
filecolor=black,%
linkcolor=apsblue,%
urlcolor=apsblue
}

\graphicspath{{plots/}:{./}}

\begin{document}

\title{Glassy dynamics and crystalline local order in two-dimensional amorphous silica}

\author{Marco Dirindin}
\affiliation{Dipartimento di Fisica, Universit\`a di Trieste, Strada Costiera 11, 34151, Trieste, Italy}

\author{Daniele Coslovich}
\email{dcoslovich@units.it}
\affiliation{Dipartimento di Fisica, Universit\`a di Trieste, Strada Costiera 11, 34151, Trieste, Italy}

\date{\today}

\begin{abstract}
We reassess the modeling of amorphous silica bilayers as a two-dimensional classical system whose particles interact with an effective pairwise potential. We show that it is possible to reparameterize the potential developed by Roy, Heyde, and Heuer to quantitatively match the structural details of the experimental samples. We then study the glassy dynamics of the reparameterized model at low temperatures.  Using appropriate cage-relative correlation functions, which suppress the effect of Mermin-Wagner fluctuations, we highlight the presence of two well-defined Arrhenius regimes separated by a narrow crossover region, which we connect to the thermodynamic anomalies and the changes in the local structure. We find that the bond-orientational order grows steadily below the crossover temperature and is associated to transient crystalline domains of nanometric size.  These findings raise fundamental questions about the nature of glass structure in two dimensions and provide guidelines to interpret the experimental data.
\end{abstract}

\maketitle

\section{Introduction}

Silicate glasses are among the first materials ever manufactured by mankind, extensively used in everyday life and technological applications.
Their disordered microscopic structure has, however, long remained a puzzle for physicists and chemists.
The early ideas of Zachariasen~\cite{Zachariasen_1932} about the glass structure have provided the foundations of the continuous random network model, which later led to the development of constraint theories~\cite{Phillips_Thorpe_1985, Gupta_Cooper_1990, Gupta_Mauro_2009}.
Modern computational studies based on classical empirical potentials~\cite{Horbach_Kob_1999, Meyer_Horbach_Kob_Kargl_Schober_2004, Carre_Berthier_Horbach_Ispas_Kob_2007}, including machine-learned ones~\cite{Liu_Fu_Li_Sabri_Bauchy_2019}, have provided direct insight into the complex network structure of amorphous silica and other silicate glasses.
It remains difficult, however, to compare subtle structural features, such as the ring statistics, with experimental data, which only provide indirect and coarse information on the local atomic arrangements.

About a decade ago, it became possible to directly visualize the microscopic structure of a two-dimensional analogue of amorphous silica.
Silica bilayers, composed by the same tetrahedral units of the 3$d$ bulk material, were in fact vapor-deposited on different kinds of substrates~\cite{lichtenstein_atomic_2012, huang_direct_2012, yu_support_2012}.
The exact mid-plane symmetry allows one to visualize the network structure using Scanning Tunneling Microscopy (STM) and Atomic Force Microscopy (AFM) techniques~\cite{buchner_two-dimensional_2017}.
When projected on a plane parallel to the substrate, the tetrahedral units formed a random network, remarkably similar to the famous Zachariasen sketches of amorphous structure~\cite{Zachariasen_1932}.
The amorphous structure has a characteristic network structure with $n$-sided rings, with $n$ ranging from 4 to 10~\cite{buchner_two-dimensional_2017}.
Thanks to these advances, it is now possible to directly compare the predictions of theoretical models of amorphous networks with experimental data~\cite{Ormrod_Morley_Thorneywork_Dullens_Wilson_2020}.

Early computational models of amorphous 2$d$ networks were based on elastic networks, but also more realistic forcefields~\cite{Wilson_Modeling_2013}.
More recently, Roy, Heyde, and Heuer (RHH)~\cite{roy_modelling_2018} developed a simple and yet realistic interaction model based on pairwise potentials, which reproduces some of the key structural features of 2$d$ silica.
The model assumes a perfect mid-plane symmetry and it is thus strictly two-dimensional.
While the radial and angular distribution functions obtained from the model agree very well with the experimental data, some discrepancies are visible in the ring statistics.
At present, it is unclear whether these are due to an intrinsic limitation of the interaction model or to insufficient experimental resolution.
It is expected that the network structure of this effective 2$d$ model bears similarities also with disordered monolayers, such as graphene or boroxine~\cite{roy_modelling_2018}. 

Two-dimensional classical systems of particles interacting via pairwise potentials have long been used as simple models to study the glass transition~\cite{Perera_Harrowell_1999, Shintani_Tanaka_2006, Kawasaki_Tanaka_2011}.
It was realized only recently, however, that the glassy dynamics in 2$d$ is severely affected by finite size effects due to the so-called Mermin-Wagner (MW) fluctuations~\cite{Flenner_Szamel_2015, Shiba_Yamada_Kawasaki_Kim_2016}.
To tackle this problem, it is customary to use cage-relative correlation functions, which measure particle motion relative to the local cage formed by the neighboring particles, as they effectively remove MW fluctuations~\cite{Shiba_Keim_Kawasaki_2018}.
Recently, Roy and Heuer~\cite{Roy_Heuer_2022} have studied the dynamics of the RHH model of 2$d$ silica in a thermal cooling process, highlighting the similarities with bulk amorphous silica, also in connection with its potential energy surface~\cite{Saksaengwijit_Reinisch_Heuer_2004}.
However, the role of MW fluctuations was not addressed in any detail.

Once expressed in terms of cage-relative observables, the qualitative features of glassy dynamics appear similar in two- and three-dimensional systems~\cite{Tarjus_2017}.
There are hints, however, that glass-forming liquids in 2$d$ may nonetheless be peculiar.
The presence of frustrated crystalline local order, which grows by decreasing temperature and has a strong connection with dynamic heterogeneity, has been underscored by numerous studies by Tanaka and collaborators~\cite{Shintani_Tanaka_2006, Kawasaki_Tanaka_2011, Tanaka_Kawasaki_Shintani_Watanabe_2010, Tong_Tanaka_2018}.
These results on 2$d$ models, however, find little correspondence in the literature on their 3$d$ counterparts~\cite{Royall_Williams_2015}, where the locally favored structure is often distinct from the one of the underlying stable crystal~\cite{Crowther_Turci_Royall_2015}.
We think that more systematic investigations on simple and efficient computational models are needed to identify robust trends across dimensionality.

In this work, we address these questions using the RHH model as a prototype of a network-forming liquid in 2$d$.
We find that a reparameterization of the RHH interaction model reproduces even the fine structural features of the experimental system, including the ring statistics.
The reparameterized model is also more efficient to simulate than the original one and enables us to study in great detail the static and dynamic correlations over a broad range of temperatures.
By removing the effect of MW fluctuations, we show that the model displays a well-defined fragile-to-strong crossover, with a low-temperature Arrhenius behavior stretching over more than $4$ decades of relaxation times.
In contrast to its 3$d$ counterpart, however, 2$d$ silica develops a pronounced crystalline local order very early during the cooling process, well before reaching the estimated melting temperature.
Our results show that 2$d$ silica is indeed a clean model to study the glass transition, but its structure displays important differences compared to its 3$d$ counterparts.

\section{Methods}
\label{sec:methods}

\begin{figure}[ht!]
\centering
\includegraphics[width=0.95\linewidth]{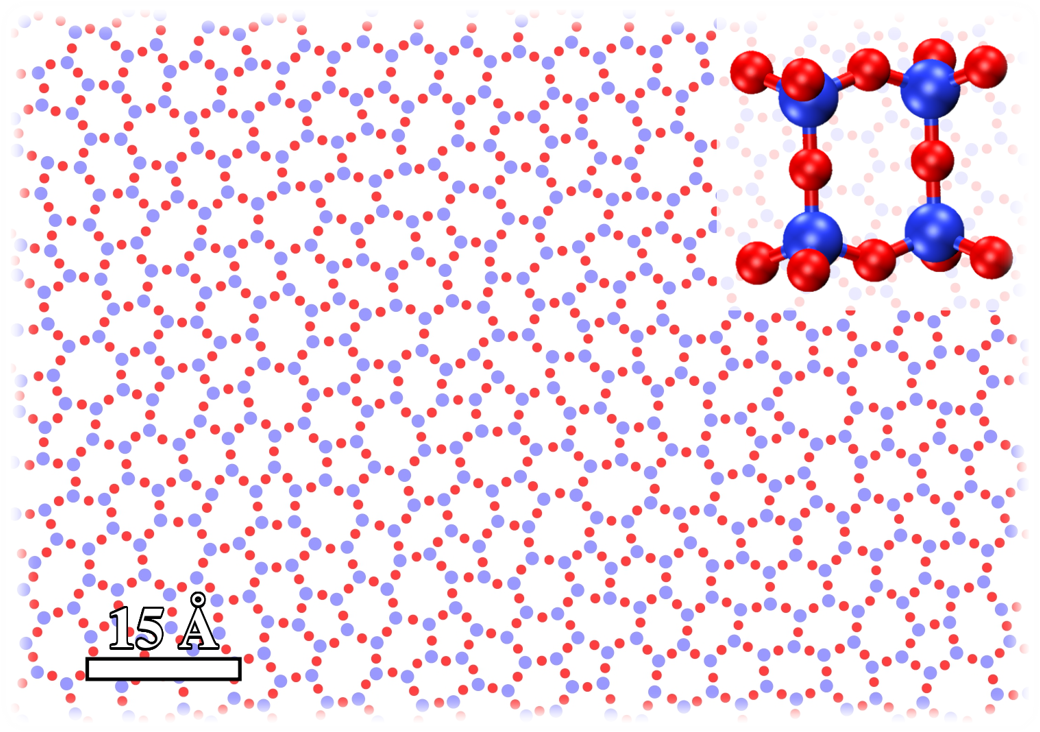}
\caption{Snapshot from the simulation of the reparameterized RHH-II model at lowest investigated temperature ($T=0.0067$). The particles form a well-defined, almost defect-free network. Blue and red circles represent effective silicon and oxygen particles, respectively. Inset: Three-dimensional rendering of the atoms in the unit cell of the silica bilayer.}
\label{fig:snap}
\end{figure}

\subsection{Models and simulation methods}
\label{sec:models}

The basic structural unit of a silica bilayer is formed by two corner-sharing $SiO_4$ tetrahedra~\cite{buchner_two-dimensional_2017}.
Since there is evidence of a perfect symmetry between the two tetrahedra with respect to the plane parallel to the substrate containing the shared oxygen~\cite{buchner_two-dimensional_2017, Wilson_Modeling_2013}, it is possible to build an effective model in which the atoms are projected on such a plane~\cite{roy_modelling_2018}.
Moreover, the substrate is not included explicitly in the model, in view of its weak van der Waals interaction with the bilayer~\cite{buchner_two-dimensional_2017, zhong_two-dimensional_2022}.
The effective system is then composed of effective particles confined on a plane, see Fig.~\ref{fig:snap}.
For simplicity, we will refer to the effective particle consisting of \textit{Si} (upper layer) - \textit{O} (intermediate layer) - \textit{Si} (lower layer) as a silicon (\textit{Si}) particle and the effective particle consisting of \textit{O} (upper layer) - \textit{O} (lower layer) as an oxygen (\textit{O}) particle.
We set the mass ratio between the effective oxygen and silicon particles to 0.57~\cite{roy_modelling_2018}.

We consider two different types of pairwise classical potentials.
The first potential features a short-range soft sphere repulsion and a long-range Yukawa interaction,
\begin{equation}
u_{\alpha \beta}(r) = \epsilon_{\alpha \beta} \left[ \left ( \frac{\sigma_{\alpha \beta}}{r} \right)^{12} + q_{\alpha \beta} \frac{e^{- \kappa r}}{r} \right] \, .
\label{eq_RHH_model}
\end{equation}
To ensure global charge neutrality, the ratio of the effective charges of \textit{Si} and \textit{O} particles is set to $Si\!:\!O\!=\!(+1.5)\!:\!(-1)$.
This model was introduced by Roy \textit{et al.}~\cite{roy_modelling_2018} and we will refer to it as the RHH model.
It captures the general features of the structure of a projected silica bilayer.
However, the ring statistics produced by the RHH model displays some discrepancies with respect to the experimental data, namely a significant fraction of large rings and a fraction of 6-sided rings appreciably lower than the experimental one.
Moreover, in the original RHH model~\cite{roy_modelling_2018}, the physical meaning of some parameters is not entirely clear: the screening parameter $\kappa$ scales with the size of the system and has no direct connection to the intrinsic length scales of the system, while the cut-off distance is set to half the side of the simulation cell, which may be overkill.
One of the aims of this work is to provide a sound basis to the model's parameters and to improve its agreement with the experiments.

The second potential we consider comprises a soft-sphere repulsion between particles of the same species and a Lennard-Jones attraction between particles of different types~\cite{coslovich_dynamics_2009} 
\begin{equation}
\begin{split}
& u_{\alpha \alpha}(r) = \epsilon_{\alpha \alpha} \left ( \frac{\sigma_{\alpha \alpha}}{r} \right)^{12} \\
& u_{\alpha \beta}(r) = 4 \epsilon_{\alpha \beta} \left [ \left ( \frac{\sigma_{\alpha \beta}}{r} \right)^{12} - \left ( \frac{\sigma_{\alpha \beta}}{r} \right)^{6}  \right ] \quad for\ \alpha \ne \beta \, .
\end{split}
\label{eq_LJ_model}
\end{equation}
We will refer to this potential as the LJ model.
This functional form was used to model 3$d$ silica~\cite{coslovich_dynamics_2009} and, despite its simplicity, reproduces some experimental properties quite well.
We also considered more complex functional forms, involving a larger number of parameters, for both kinds of models.
For example, we considered different exponent powers and different screening parameters for each type of interaction.
However, we did not find any major improvements over the models presented above. 
In the following, we will use reduced units, assuming $r_{unit}=1$\ \AA\ and $m_{unit}=72\ g/mol$~\cite{roy_modelling_2018} as the units of length and mass, respectively, while we set the unit of energy $e_{unit}$ equal to $\epsilon_{SiSi}$ for each potential.
We define the time unit as
$t_{unit} = r_{unit} (\frac{m_{unit}}{e_{unit}})^{1/2}$.

We consider systems composed of $N$ particles enclosed in a rectangular cell of sides ($L_x$, $L_y$) with periodic boundary conditions.
We fixed the number density at the experimental value, $\rho=N / (L_x L_y)=0.2068$~\cite{roy_modelling_2018}.
We set the ratio $L_y/L_x$ so as to yield a nearly square cell commensurate with the honeycomb crystal lattice at the chosen chemical composition $x_{Si}=2/5$.
We carried out molecular dynamics simulations using \texttt{RUMD}~\cite{bailey_rumd_2017}, which is particularly efficient for the system sizes of interest, which range from $N=80$ to $10^5$ particles.
We performed all our simulations in the NVT ensemble using a Nose-Hoover thermostat with a thermostat relaxation time $\tau_{term}=1.6381$.
The integration timestep was set to 0.045 at $T=0.015$ and then scaled with the square root of the inverse temperature.

We performed two kinds of simulations: the first ones to produce out-of-equilibrium glass states and the second ones to simulate liquids at equilibrium conditions.
In the first case, the quench protocol consisted in a sequence of simulations at temperatures 0.0200, 0.0125, 0.0100, 0.0075, and 0.0010.
These simulations lasted 50000, 50000, 250000, 250000, and 5000 timesteps, respectively.
Note that this protocol does not correspond to the experimental one, which is based instead on physical vapor deposition and can be optimized to produce even very stable glasses~\cite{Swallen_Kearns_Mapes_Kim_McMahon_Ediger_Wu_Yu_Satija_2007}.
Nonetheless, we will show in Sec.~\ref{sec:liquid:structure} that the structure of these rapidly quenched, out-of-equilibrium states can match very well the experimental data.
For the equilibrium simulations, we start from a configuration sampled at high temperature and then equilibrate the system at the desired temperature. We anneal the system until the mean square displacement (MSD) of particles of both species exceeds $10^3$. We then perform a production run at the same temperature to store configurations, making sure again that the number of steps is such that the final MSD is always at least $10^3$.

\subsection{Structure-matching optimization}
\label{sec:structure_matching}

To determine the interaction parameters, we used a structure-matching approach. 
After some trial and errors, we decided to fit only a limited set of structural features, namely the average nearest-neighbor distance between effective silicons and the experimental ring statistics.
We obtained the nearest-neighbor distance from the position of the first peak of the radial distribution function (RDF) between silicons, $g_{SiSi}(r)$.
Our definition of ring is based on graph theory and is discussed in detail in the Appendix~\ref{sec:appendix:rings}.
In essence, we consider the network of silicon particles and we identify the rings with the faces of that graph.
The ring statistics is then defined as the probability distribution $P(n)$ of rings with a given number of sides $n$.
Empirically, we found that fitting the \textit{Si-Si} distance and the rings statistics is enough to reproduce a larger set of structural features, including the full shape of the RDFs and the bond-angle distributions, see Sec.~\ref{sec:rdf_bad}. 

With that in mind, we aim at minimizing the cost function
\begin{equation}
\chi^2(\mathbf{q}) = \chi_{RDF}^2(\mathbf{q}) + \chi_{ring}^2(\mathbf{q}) + \chi_{pen.}^2(\mathbf{q})  \, ,
\end{equation}
where $\mathbf{q}$ is the vector of interaction parameters and
\begin{align*}
\chi_{RDF}^2(\mathbf{q}) = \ & \omega_{RDF}^{SiSi} \left[ r_{SiSi}(\mathbf{q}) - \tilde{r}_{SiSi}\right]^2 \, , \\
\chi_{ring}^2(\mathbf{q}) = \ & \sum_{n=4}^{10} \omega_{ring}^{n}\left[ P(n;\mathbf{q}) - \tilde{P}(n) \right]^2 \, , \\
\chi_{pen.}^2(\mathbf{q}) = \ & \omega_{pen.}^3 \, P(3;\mathbf{q})^2 + \omega_{pen.}^{n>10} \, P(n>10;\mathbf{q})^2 \, .
\end{align*}
Here, $r_{SiSi}$ is the position of the first maximum in the \textit{Si-Si} RDF, $\omega_i^j$ are weights, and the quantities with a tilde on top are those measured experimentally.
In particular, $\omega_{RDF}^{SiSi} = 1$, $\omega_{ring}^{n} = 3.629025 \cdot 10^{-4} / \sigma_n^2$ with $\sigma_n$ the standard deviation on $\tilde{P}(n)$, 
and $\omega_{pen.}^3=\omega_{pen.}^{n>10}=45$.
We obtained the experimental RDF from the configurations in Refs.~\cite{lichtenstein_atomic_2012-1, lichtenstein_crystalline-vitreous_2012}.
We determined the experimental ring statistics by averaging over 5 samples synthesized by different groups on different substrates~\cite{kumar_ring_2014}, see the Appendix~\ref{sec:appendix:exp:samples}.
Since these samples show very similar ring statistics, we decided to include all of them in the evaluation of the cost function.
The cost function also includes a penalty term, $\chi_{pen.}^2$, that prevents a too large fraction of small and large rings, since rings of this type are not observed experimentally.
Note that all the experimental quantities involved in $\chi^2(\mathbf{q})$ are measured at low temperature, due to constraints on the STM and TEM techniques used to obtain the atomic positions.

\begin{figure}[!tb]
\centering
\includegraphics[width=0.95\linewidth]{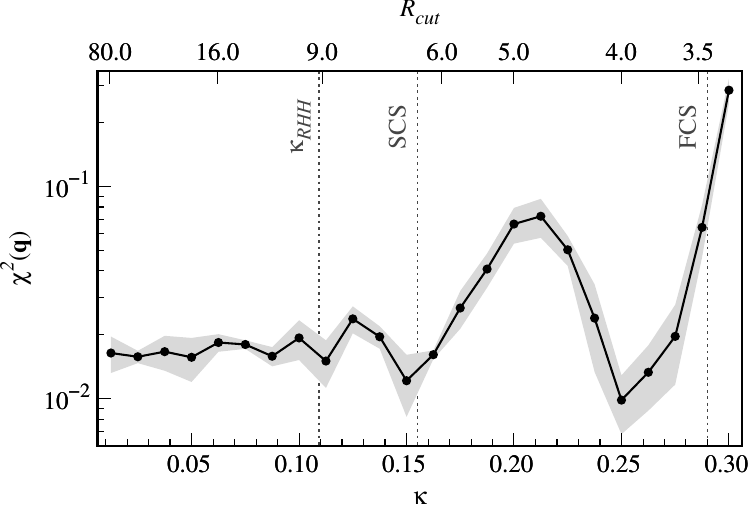}
\caption{Dependence of the cost function $\chi^2$ on the screening parameter $\kappa$ for the original RHH model. The cutoff parameter was fixed to $R_{cut} = 1/\kappa$. The shaded area is an estimate of the statitical uncertainty. The vertical dashed lines indicate the first coordination shell (FCS) and the second coordination shell (SCS) of $g_{SiSi}(r)$ and the original values of the screening parameter and the cutoff radius used by Roy \textit{et al.}~\cite{roy_modelling_2018} ($\kappa_{RHH}$).}
\label{img_RHH_cost_vs_k}
\end{figure}

Some comments about the parameters of the RHH model are in order.
In the original formulation of the RHH model, both the cutoff radius and the screening length $l_C = \frac{1}{\kappa}$ scale linearly with the cell side~\cite{roy_modelling_2018}.
This results in long-range interactions that become progressively more expensive to compute as the system size increases.
To determine the optimal values of these parameters, we proceeded as follows.
Since the cost function depends weakly on both $\kappa$ and $R_{cut}$, we optimized them separately: we evaluated the cost function changing one of these parameters at the time, while fixing the others to their values in the original RHH model.
We found empirically that fixing $R_{cut} = \frac{1}{\kappa}$ produced overall good results,
hence we optimized the screening parameter using this constraint.
As shown in Fig.~\ref{img_RHH_cost_vs_k}, the cost function becomes nearly constant when $\kappa$ is smaller than 0.1, close to the value used by RHH for $N=80$ ($\kappa_{RHH}=0.1093$).
We also found that the cost function has two minima, corresponding to values of $l_C$ just after the first coordination shell and around the second coordination shell of $g_{SiSi}(r)$, respectively.
These results indicate that the network structure depends mainly on interactions up to the second shell of neighbors.
Using this information, we optimized the parameters for the RHH model fixing $\kappa=\frac{1}{l_C}$ and $R_{cut}$ just after the first (4.0) or the second coordination shell (6.0).
We will call the corresponding reparameterized models RHH-I and RHH-II, respectively.
Finally, we also fixed $\epsilon_{\alpha \beta} = 1$ for the RHH model and $\epsilon_{\alpha \alpha} = 1$ for the LJ model, because we did not find any significant decrease in the cost function when these parameters are varied.
The only parameters to optimize are then $\mathbf{q} = \{ \sigma_{SiSi}, \sigma_{SiO}, \sigma_{SiO} \}$ for the RHH model and $\mathbf{q} = \{ \sigma_{SiSi}, \sigma_{SiO}, \sigma_{SiO}, \epsilon_{SiO} \}$ for the LJ model.

The evaluation of the cost function for a given $\mathbf{q}$ was carried out as follows.
We started by simulating the original RHH model at high temperature ($T=0.0200$), well within the liquid phase, and stored $N_{conf}$ independent configurations.
Starting from each of these configurations, the system was cooled down using the procedure to produce glass states described in Sec.~\ref{sec:models}.
We then computed the RDFs and the ring statistics using all the $N_{conf}$ final configurations we obtained.

To minimize the cost function, we used the Levenberg-Marquardt algorithm implemented in \texttt{lmfit}~\cite{newville_lmfit_2014}.
For the minimization, we used systems of 500 particles and we averaged the ring statistics over $N_{conf}=10$ configurations.
Once a minimum was found, we carried out ten additional minimizations starting from a small random perturbations of the parameters found in the minimum.
Then, we compared the value of the cost functions obtained for this set of minima.
For this evaluation, we increased the system size to 7000 particles and we averaged over $N_{conf}=50$ configurations.
We selected the optimal model as the set of parameters with the lowest $\chi^2$ among these refined minimizations.
The optimized parameters of each model are summarized in Table~\ref{tab_parameters}

\begingroup
\setlength{\tabcolsep}{4pt}
\begin{table}[ht!]
\centering
\begin{tabular}{c l l l c c c}
\toprule
\textbf{Model} & \textbf{$\sigma$}$_{SiSi}$ & \textbf{$\sigma$}$_{SiO}$ & \textbf{$\sigma$}$_{OO}$ & \textbf{$\epsilon$}$_{SiO}$ & \textbf{$\kappa$} & \textbf{$R_{cut}$} \\
\midrule
\midrule
\textbf{RHH} 		& 2.25		& 1.075	& 0.900	& $-$		& 0.1093	& 9.1518						\\
\textbf{RHH-II}	& 2.3015	& 1.0873	& 0.9227	& $-$		& 0.1675	& 6.00							\\
\textbf{RHH-I~}	& 2.3039	& 1.0611	& 0.8899	& $-$		& 0.2500	& 4.00							\\
\textbf{LJ}		& 2.3445	& 1.3582	& 1.6913	& 0.1708	& $-$		& $2.5\cdot \sigma_{\alpha\beta}$	\\
\bottomrule
\end{tabular}
\caption{Interaction parameters of the models studied in this work. Note that the values of $\kappa$ e $R_{cut}$ for the RHH model are those used in Ref.~\cite{roy_modelling_2018} for $N=80$ particles.}
\label{tab_parameters}
\end{table}
\endgroup

\section{Results: comparison with the experimental glass structure}

In this section, we assess the ability of the reparameterized potentials to reproduce the structure of 2$d$ silica observed experimentally.
We use glass configurations obtained from the simulation protocol described in Sec.~\ref{sec:models}, whereby a high-temperature liquid is rapidly cooled down to nearly zero temperature.
We examine both simple structural quantities, such as two and three-body correlation functions (Sec.~\ref{sec:rdf_bad}), and finer indicators of the network structure, such as the rings statistics and ring correlations (Sec.~\ref{sec:rings}).
Finally, we compare the computational efficiency of the new potentials to that of the original RHH model (Sec.~\ref{sec:exec}).

\subsection{Radial distribution functions and bond angle distributions}
\label{sec:rdf_bad}
We start by analyzing some conventional structural quantities: the partial RDF and the bond-angle distribution (BAD) functions.

\begin{figure}[!tb]
\centering
\includegraphics[width=0.95\linewidth]{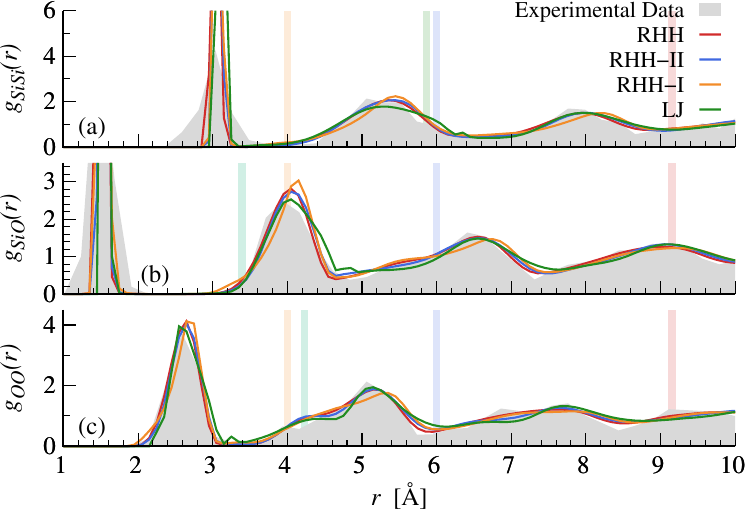}
\caption{Comparison between the experimental radial distribution functions and the ones of the models studied in this work: (a) $g_{SiSi}(r)$, (b) $g_{SiO}(r)$, and (c) $g_{OO}(r)$. The experimental data refer to a silica bilayer on Ru(0001), see the  Appendix~\ref{sec:appendix:exp:samples}. The vertical bands mark the cutoff radii of the corresponding pair potentials for the RHH (red), RHH-II (blue), RHH-I (orange), and LJ (green) models.}
\label{img_rdf}
\end{figure}

In~Fig.~\ref{img_rdf}, we show the partial RDF $g_{\alpha\beta}(r)$, obtained from the experimental samples and from the models introduced in Sec.~\ref{sec:methods}.
All the models reproduce the experimental RDFs pretty well.
Note that, unlike Roy \textit{et al.}~\cite{roy_modelling_2018}, we did not convolve the RFDs with a Gaussian function to emulate the experimental uncertainty on the measured positions.
The first peaks of $g_{SiSi}(r)$ and $g_{SiO}(r)$ obtained from our models are thus higher than observed experimentally, but we do not deem this difference relevant to our analysis.
Indeed, the peaks' positions are always correctly reproduced by our models.

\begin{figure}[!tb]
\centering
\includegraphics[width=0.95\linewidth]{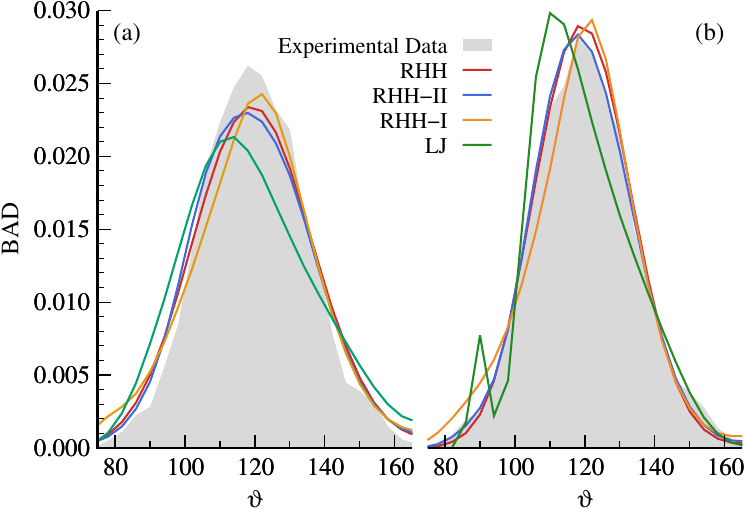}
\caption{Comparison between the experimental bond-angle distributions and the ones of the models studied in this work for (a) \textit{Si-Si-Si} (left) and (b) \textit{O-Si-O} angles (right). The experimental data refer to a silica bilayer on Ru(0001), see the Appendix~\ref{sec:appendix:exp:samples}.}
\label{img_bond_angle_distributions}
\end{figure}

To better evaluate the ability of the interaction models to reproduce the experimental structure, we analyze the BADs, which measure the distribution of angles produced by three neighboring particles.
The neighbor distance between particles of species $\alpha$ and $\beta$ is defined as the position of the first minima in the corresponding partial $g_{\alpha\beta}(r)$ and corresponds to 3.90, 2.60 e 3.50 for \textit{Si-Si}, \textit{Si-O} and \textit{O-O} neighbors, respectively.
In~Fig.~\ref{img_bond_angle_distributions}, we compare the experimental and simulated BADs for selected triplets of neighboring species.
We find that the RHH model and its variants capture the essential features of the experimental system: both the distribution of \textit{Si-Si-Si} and \textit{O-Si-O} angles are peaked at 120$^\circ$ and are symmetrical about the maximum.
The LJ model is instead visibly worse than the others and reproduces these characteristics only approximately.
In particular, one notices a spurious peak around 90$^\circ$, possibly due to an excess of square local arrangements. From Fig.~\ref{img_bond_angle_distributions}, however, we find no clear difference between the RHHs models, and we can only conclude that the LJ model performs poorly compared to the potentials with a RHH functional form.

\subsection{Ring statistics and correlations}
\label{sec:rings}

Since traditional structural indicators did not reveal significant differences between the RHH models, we turn to many-body properties related to the network structure.
We start by analyzing the probability distribution $P(n)$ of rings with a given number of sides $n$.
Following the convention for silica systems without homopolar bonds~\cite{le_roux_ring_2010}, we define the number of sides of a ring as the number of silicon atoms it contains.

\begin{figure}[!tb]
\centering
\includegraphics[width=0.95\linewidth]{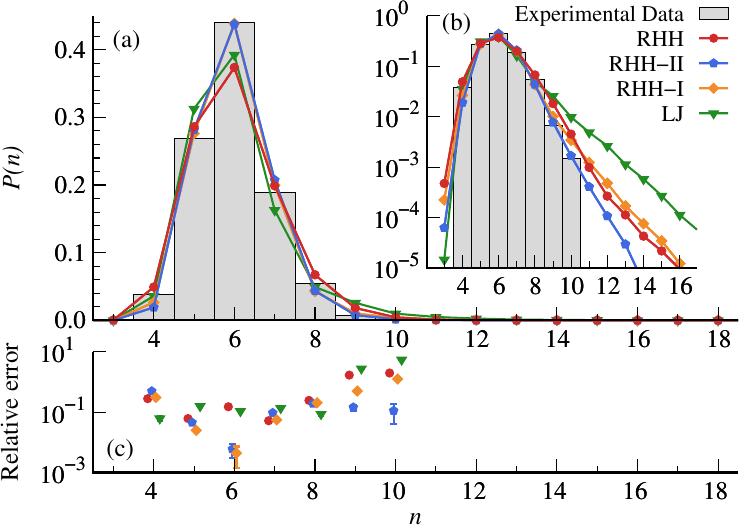}
\caption{Comparison between the experimental ring statistics and the ones of the models studied in this work, in (a) linear and (b) semi-logarithmic scales. Note that in panel (a) the data for the RHH-I and RHH-II models are almost superposed on top of each other.
  (c) Relative errors between the experimental ring statistics and the one of each studied model. The experimental data are the average over several samples grown on different substrates, see the Appendix~\ref{sec:appendix:exp:samples}. The error bars are shown only when larger than the symbol size.} 
\label{img_ring_statistics}
\end{figure}

As the ring statistics was directly targeted by our structure-matching procedure, we expect an improved agreement with the experimental data for the reparameterized models compared to the original RHH model.
From Fig.~\ref{img_ring_statistics}, we see that both RHH-I and RHH-II models improve significantly the agreement between experimental and simulated ring statistics.
In particular, they accurately reproduce the high fraction of hexagonal rings, which is underestimated by the standard RHH model~\cite{roy_modelling_2018}.
The RHH-II model also reduces the fraction of very large rings, i.e. rings with more than 10 sides, as it can be appreciated from the inset of Fig.~\ref{img_ring_statistics}.
Due to the much higher statistics available in the simulations, the models always display an exponential tail of very large rings, which can be assumed to reflect a Poisson distribution.
We note again that the LJ model is unable to correctly reproduce the experimental ring statistics.
In particular, it does not produce the high fraction of hexagonal rings and has a very fat tail at large $n$.

\begin{figure}[!tb]
\centering
\includegraphics[width=0.95\linewidth]{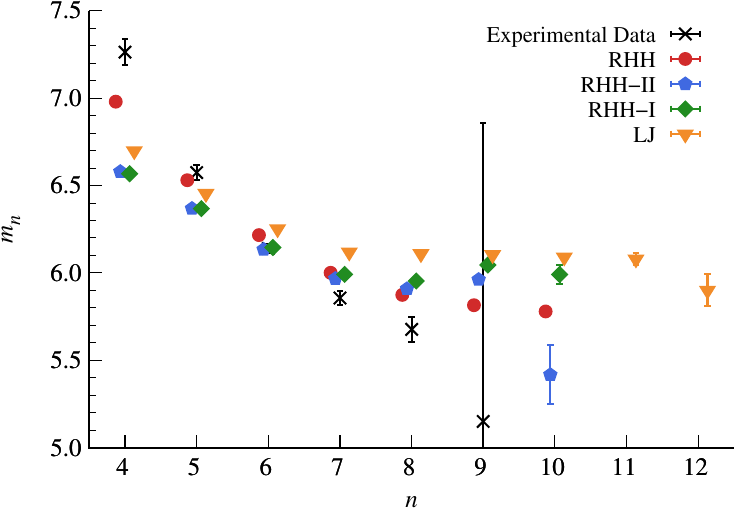}
\caption{Relationship between the number of sides $n$ of a ring and the average number of sides $m_n$ of its neighbors. A clear anticorrelation between these quantities can be observed for both the experimental data and the models studied in this work. The experimental data are calculated ignoring the rings on the boundaries of the configuration, see the Appendix~\ref{sec:appendix:exp:samples}.}
\label{img_AW_law}
\end{figure}

The differences among the models are further highlighted by studying correlations between neighboring rings.
In the networks formed by a large number of natural systems, such as polycristalline grains~\cite{aboav_arrangement_1970}, two-dimensional glasses~\cite{kumar_ring_2014, roy_modelling_2018, ebrahem_vitreous_2020}, foams~\cite{weaire_soap_1984} and others, large rings tend to be surrounded by smaller ones~\cite{Ormrod_Morley_Thorneywork_Dullens_Wilson_2020}.
This anti-correlation is underscored by the empirical Aboav-Weaire law, which affirms that $m_n \sim A  + \frac{B}{n}$, where $m_n$ is the average ring size of the neighbors of a ring of size $n$, while $A$ and $B$ are fitting parameters.
Despite its remarkable generality, it is difficult to compare different systems using the Aboav-Weaire law and its parameters do not have a clear connection with the network structure~\cite{Ormrod_Morley_Thorneywork_Dullens_Wilson_2020}.
Here, we quantify ring correlations by showing $m_n$ as function of $n$, see Fig.~\ref{img_AW_law}.
This representation clearly shows that larger rings tend to be connected to smaller ones, unless $n$ is large enough that differences in the local network environment are averaged out.
Note that, due to the small sample sizes, the saturation of $m_n$ at large $n$ cannot be appreciated from the experimental data, while it is evident from the simulations.

A standard measure of the correlation between rings in network theory is the assortativity coefficient, 
which is a quantity in the range $[-1, +1]$ that measures the likelihood that two connected nodes of the network have similar properties~\cite{newman_networks_2018}.
In particular, this coefficient is $+1$ ($-1$) if the network has assortative (dissortative) mixing, while it is 0 if the node properties are randomly distributed among the nodes.
Here, assortative mixing means that nodes with similar properties tend to be connected together, while dissortative mixing means that the properties of linked nodes have opposite fluctuations with respect to the mean.

We computed the assortativity coefficient by considering a network where the rings are the nodes and two nodes are joined if the corresponding rings share at least one edge.
As a node property, we use the number of sides, $n$, of the ring.
The assortativity coefficients obtained for the different models are compared in  Table~\ref{tab_assortativity}.
All the models have a negative assortativity coefficient, meaning that the ring structures produced by our models have small rings close to large ones, like the experimental systems.
Although the assortativity coefficients are all negative, we observe a general trend in their magnitude: the longer the range of the potential, the more dissortative is the network.
However, within statistical uncertainties, both the original and the reparameterized RHH models produce dissortative network structures compatible with experimental data.
Interestingly, the value we obtained for the assortativity coefficient of the experimental system is appreciably lower than the one reported in Ref.~\cite{Ormrod_Morley_Thorneywork_Dullens_Wilson_2020} and is not compatible with it within error bars.
Our findings indicate that the agreement between the networks produced by the raft algorithm and the experimental network structure of 2$d$ silica, observed in Ref.~\cite{Ormrod_Morley_Thorneywork_Dullens_Wilson_2020}, may be coincidental.

\begingroup
\setlength{\tabcolsep}{4pt}
\begin{table}[!tb]
\centering
\begin{tabular}{c c c}
\toprule
\multirow{2}{4em}{\textbf{Model}} 	& \textbf{Assortativity}	& \textbf{Interaction} 	\\
											& \textbf{coefficient}		& \textbf{range}	\\
\midrule
\midrule
\textbf{Experiment}	 	& $-0.22\quad \pm 0.06\quad$ 		& $-$	 			\\
\textbf{RHH} 				& $-0.2185 \pm 0.0003$ 			& $7.4$ 	 		\\
\textbf{RHH-II}			& $-0.1610 \pm 0.0004$ 			& $5.1$ 	 		\\
\textbf{RHH-I~}			& $-0.1323 \pm 0.0005$ 			& $3.6$ 		 	\\
\textbf{LJ}				& $-0.0876 \pm 0.0006$ 			& $2.9$ 		 	\\
\bottomrule
\end{tabular}
\caption{Comparison between the assortativity coefficients of the models studied in this work. The interaction range is calculated as the distance at which the potential energy between a silicon and an oxygen particle is 1 \% of its value at the minimum.}
\label{tab_assortativity}
\end{table}
\endgroup

\subsection{Overall assessment and computational efficiency}
\label{sec:exec}
Our structural analysis of the glass states shows that the RHH-II model provides the best overall agreement with the experimental system.
It correctly reproduce the behavior of the RDFs, the BADs and the rings statistics. 
Its network structure also possesses a marked dissortative mixing in rings sizes, as observed experimentally.
Quantitatively, the RHH-II model produces the lowest cost function~($\chi^2$), improves the agreement with the experimental ring statistics compared to the standard model~($\chi_{ring}^2$) and has the smallest fraction of very large rings~($\chi_{pen.}^2$), see Table~\ref{tab_conclusion}.

One of our goals in reparameterizing the RHH model was also to improve its computational efficiency.
When using \texttt{RUMD}, we found that the RHH-II model is around 60\% more efficient than the original RHH model, due the smaller range of the potential.
The efficiency ratio we observe for both reparameterized RHH models is, however, lower than what we can expect from the number of interacting particles computed from the RDFs.
Given the small number of interacting particles in the reparameterized models, this effect may be due to GPU undersaturation, which makes the computation inefficient from the hardware perspective.
Still, the speed-up observed in a direct evaluation of the forces is about 2.28, which may be valuable when considering the choice of the model.

\begingroup
\setlength{\tabcolsep}{4pt}
\begin{table}[!tb]
\centering
\begin{tabular}{c c c c c c c}
\toprule
\multirow{2}{2em}{\textbf{Model}} & \multirow{2}{2em}{\textbf{$\chi^2$}} & \multirow{2}{2em}{\textbf{$\chi_{RDF}^2$}} & \multirow{2}{2em}{\textbf{$\chi^2_{ring}$}} & \multirow{2}{2em}{\textbf{$\chi^2_{pen.}$}} & \multicolumn{2}{c}{\textbf{Efficiency Ratio}} \\
 & & & & & Forces & MD \\
\midrule
\midrule
\textbf{RHH} 		& \hphantom{0}1.47 		 	& \textbf{0.03}	& \hphantom{0}0.97 			& 0.46			& 1.00 		& 1.00 \\
\textbf{RHH-II}	& \hphantom{0}\textbf{0.32}	& \textbf{0.03}	& \hphantom{0}0.24 			& \textbf{0.04}	& 2.28			& 1.64 \\
\textbf{RHH-I~}	& \hphantom{0}1.24 		  	& \textbf{0.03}	& \hphantom{0}\textbf{0.19}  	& 1.01			& 2.52			& 2.12 \\
\textbf{LJ}		& 21.24 			& 1.41				& 2.67 			& 17.17				& \textbf{2.88} & \textbf{2.88} \\
\bottomrule
\end{tabular}
\caption{Comparison between the original RHH model and the ones studied in this work, see text for definitions. The best values for each indicator are shown in bold.}
\label{tab_conclusion}
\end{table}
\endgroup

\section{Results: glassy dynamics and structure}

In this section, we turn our attention to the equilibrium simulations of 2$d$ liquid silica, modeled by the RHH-II potential.
Our simulations span a broad range of temperatures, from the normal liquid down to highly viscous states.
We start by characterizing the thermodynamics and network structure (Sec.~\ref{sec:liquid:structure}).
We then show that the system develops a strong local crystalline order upon cooling, while remaining stable against (quasi) long range order (Sec.~\ref{sec:liquid:T}). Finally, we analyze the dynamics of the liquid at equilibrium, paying particular attention to removing Mermin-Wagner fluctuations (Sec.~\ref{sec:liquid:MW}) and identifying distinct dynamic regimes (Sec.~\ref{sec:liquid:dynamics}).

\begin{figure}[!ht]
\centering
\includegraphics[width=0.95\linewidth]{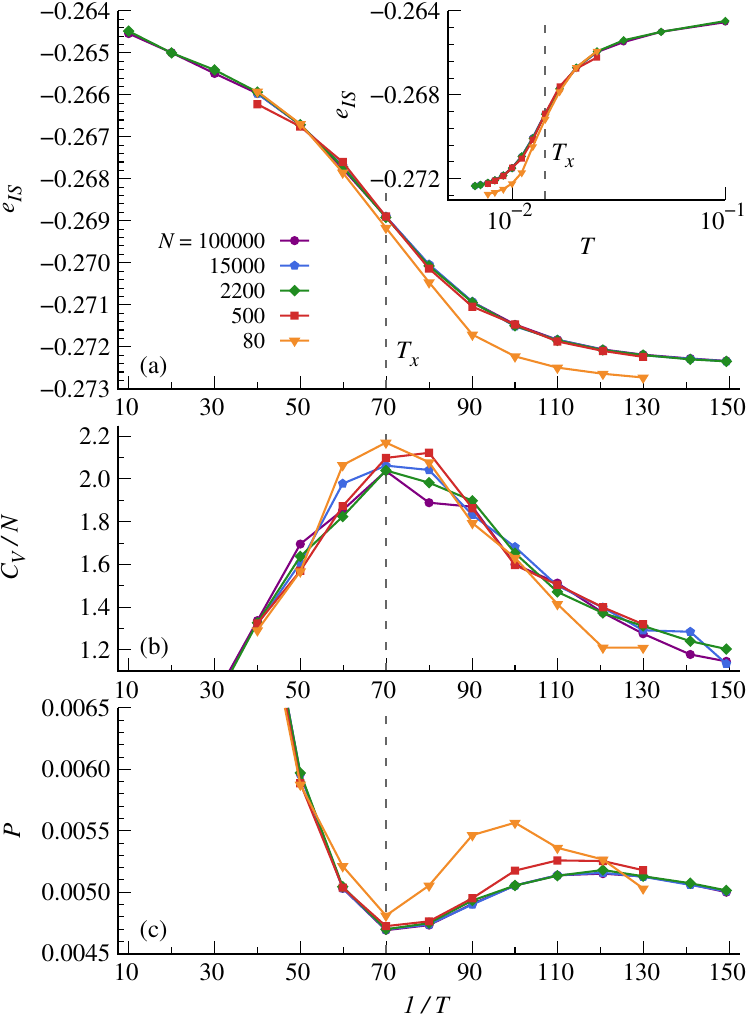}
\caption{Temperature dependence of (a) the average potential energy per particle $e_{IS}$ of the inherent structures, (b) excess specific heat per particle $C_V/N$, and (c) pressure $P$, for different system sizes, as indicated in the figure. The inset of panel (a) shows $e_{IS}$ versus $T$ in semi-logarithmic scale.}
\label{fig:thermo}
\end{figure}

\subsection{Thermodynamic anomalies and network structure}
\label{sec:liquid:structure}

It is well-known that the formation of a random network structure in bulk 3$d$ silica is accompanied by thermodynamic anomalies, such as a minimum in pressure (at constant density) and a peak in the specific heat~\cite{Horbach_Kob_1999, Saika-Voivod_Poole_Sciortino_2001, Saika-Voivod_Sciortino_Poole_2004}.
These anomalous features are common to other network-forming liquids, such as silicon and water, and have been attributed to polyamorphism, i.e., the presence of different liquid phases, as well as to changes in the properties of the potential energy landscape~\cite{Tanaka_2022, Heuer_2008}.

To check whether 2$d$ silica follows similar trends, we collect in Fig.~\ref{fig:thermo} several thermodynamic indicators as a function of temperature.
We first inspect the energy $e_{IS}(T)$ of the underlying inherent structures, i.e., the local minima determined by a local minimization of the potential energy from equilibrium configurations sampled at a given temperature~\cite{Stillinger_Weber_1982}.
From Fig.~\ref{fig:thermo}(a) we see that $e_{IS}$ is approximately constant at high temperature and drops markedly below $T\approx 0.02$.
As found in other 3$d$ glass-forming liquids~\cite{Sastry_Debenedetti_Stillinger_1998}, this crossover temperature also marks the onset of glassy dynamics, see Sec.~\ref{sec:liquid:MW}.

Systems with a Gaussian statistics of inherent structures display a simple scaling $e_{IS} \sim 1/T$ below the onset temperature~\cite{Heuer_Buchner_2000}.
While models of fragile glass-forming liquids are typically well-described by such a Gaussian statistics, 
in 3$d$ silica this regime is very narrow, if existent at all, and $e_{IS}$ saturates rapidly at low temperatures, going through an inflection point.
Our results for the RHH-II model confirm this result, see also Ref.~\cite{Roy_Heuer_2022}.
We see no evidence of an extended linear regime, $e_{IS}\sim 1/T$, in Fig.~\ref{fig:thermo}(a).
The inflection point occurs at a crossover temperature $T_x\approx 0.014$.
The saturation of $e_{IS}$ below $T_x$ has been interpreted as the approach to the ``bottom of the landscape'', as evidenced by the presence of a lower bound to the inherent structure energy distribution~\cite{Roy_Heuer_2022}, and has been connected to a fragile-to-strong crossover~\cite{Saksaengwijit_Reinisch_Heuer_2004, Roy_Heuer_2022}, see Sec.~\ref{sec:liquid:dynamics}.

Turning to conventional thermodynamic quantities, Fig.~\ref{fig:thermo}(b) and (c) show the excess specific heat per particle $C_V/N=\left. \frac{dU}{dT} \right|_V$ and the pressure $P$, respectively.
Results for different system sizes highlight the lack of major finite size-effects on thermodynamic quantities, except for the smallest system size ($N=80$).
The specific heat displays a broad peak around $T_x\approx 0.014$, and the pressure has a minimum at the same temperature.
These features appear qualitatively similar to those found at constant density in 3$d$ amorphous silica modeled with the BKS potential~\cite{Saika-Voivod_Sciortino_Poole_2004}.
Note that both anomalies occur at the same crossover temperature $T_x$ in the RHH-II model, while a clear discrepancy between the two is found in BKS silica.

\begin{figure*}[!ht]
\centering
\includegraphics[width=0.95\textwidth]{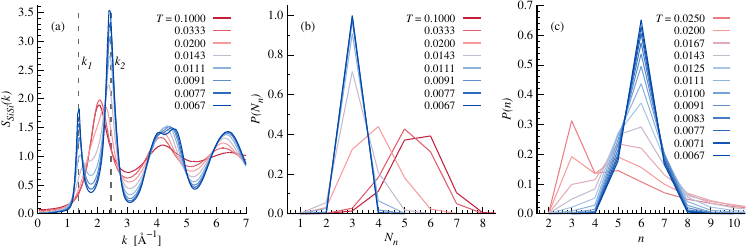}
\caption{(a) Structure factor $S_{SiSi}(k)$, (b) distribution of the number of \textit{Si-Si} neighbors and (c) ring statistics $P(n)$ at different temperatures. The structure factors and the neighbors distributions are calculated for a system with $10^5$ particles, while the ring statistics are calculated for a system with 15000 particles.}
\label{fig:structure}
\end{figure*}

To connect the thermodynamic anomalies shown in Fig.~\ref{fig:thermo} to the evolution of the liquid structure, we show in Fig.~\ref{fig:structure}(a) the structure factor $S_{SiSi}(k)$ between silicon atoms.
At high temperature, the $S_{SiSi}(k)$ displays a sharp peak, centered around $k\approx 2$, which splits upon cooling into a ``first sharp diffraction peak'' at $k_1=1.375$ and a main peak at $k_2=2.475$.
These wave-numbers corresponds to the typical length scales of the rings and of the \textit{Si-Si} bonds, respectively, similar to what found in 3$d$ silica.
Upon cooling, the liquid structure thus crosses over from the one typical of a ``normal'' liquid to the one of a network liquid.
To pinpoint this crossover, we look for an inflection of $S_{SiSi}(k)$ in the range between $k_1$ and $k_2$.
Remarkably, it occurs around the same crossover temperature $T_x$ identified from the thermodynamic analysis.
For comparison, in BKS silica the inflection of $S_{SiSi}(k)$ occurs around $T=6000$ K~\cite{Horbach_Kob_1999}, while the peak of $C_V$ around $T=4800$ K~\cite{Saika-Voivod_Poole_Sciortino_2001}.

The formation of a network structure is demonstrated by the probability distributions of the number of \textit{Si}-neighbors of \textit{Si} particles and of the ring sides $n$.
These are shown in Fig.~\ref{fig:structure}(b) and (c), respectively.
Above $T_x$, the network is not well-defined, as it can be seen from the presence of spurious 3-fold rings and the broad distribution of number of neighbors.
Below $T_x$, instead, the liquid develops a network structure with the characteristic features of 2$d$ amorphous silica, \textit{i.e.}, 3-fold local arrangements and 6-membered rings are the locally favored structures of the network structure.
By comparing these results to the experimental data used to match the structure of the RHH-II glass, we observe that the former resembles a liquid equilibrated at around $T \approx 0.0100$, while our lowest equilibrated temperature is about twice lower than that.
This shows that the experimental amorphous samples of 2$d$ silica are representative of poorly annealed glass states.

\subsection{Local crystalline order and melting temperature}
\label{sec:liquid:T}

The locally favored structure found in the low-temperature liquid is an equilateral triangle centered around $Si$ particles, thus sharing the same local symmetry as the crystal~\footnote{Note that, according to the Mermin-Wagner theorem~\cite{Mermin_1968}, the crystalline phase can only display quasi-long range order}.
It is natural, therefore, to ask whether these structures develop some orientational order beyond the first coordination shell and whether there is any interplay with crystallization.
This kind of analysis may also shed some light on the structure of silica bilayers observed in experiments, which show some local regions with pronounced crystalline order~\cite{huang_direct_2012, lichtenstein_crystalline-vitreous_2012, buchner_two-dimensional_2017}.

To characterize the bond-orientational order, it is natural to introduce a set of order parameters that quantify the local alignment of bonds with $n$-fold local symmetry.
We define the complex bond-order parameter for an $n$-fold symmetry of the $j$-th particle as
\begin{equation}
  \label{eqn:psi_n}
\psi_{n,j}(t) = \frac{1}{n} \sum_{l=1}^n e^{im\phi_{jl}(t)}  \, ,
\end{equation}
where $\phi_{jl}(t)$ is the angle between the bond that connects particles $j$ and $l$ at time $t$ and an arbitrary reference line, and the sum runs over the first $n$ nearest neighbors of $j$.
In this section, we will focus on the \textit{Si} particles as central particles for the calculation of $\psi_{n,j}$. Due to the $3$-fold local antiferromagnetic symmetry around \textit{Si} particles, only $n=3$ is relevant for correlations within the first coordination shell.
We also define a coarse-grained local order parameter by averaging over the first $Si$-$Si$ coordination shell, $\psi^{cg}_{n,i}=\frac{1}{3} \sum_{j=1}^3 \psi_{3,j}$, and the bond-order parameter $\psi^R_{6,j}$ between the rings, using the centers of mass of the \textit{Si} rings as effective coordinates in Eq.~\eqref{eqn:psi_n}.

\begin{figure*}[!tb]
\centering
\includegraphics[width=0.95\textwidth]{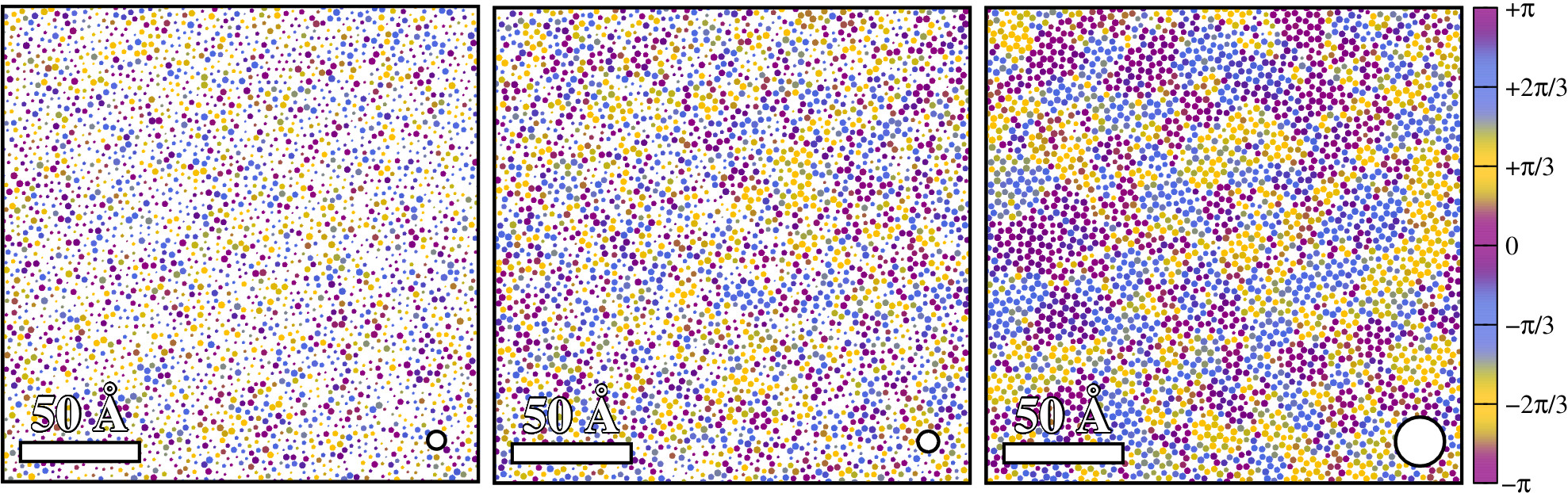}
\caption{Snapshots illustrating the development of local crystalline order. The particles are color-coded according to the phase of the complex local order parameter $\psi_{3,i}$ between \textit{Si-Si} atoms, while their size is proportional to $|\psi_{3,i}|$. The configurations are taken at temperature $T=0.0250$, $0.0143$ ($\approx T_x$), and $0.0067$ (from left to right). The white circle in the lower right corner of each configuration has a radius equal to the correlation length at the corresponding temperature. The palette features repeating colors due to the antiferromagnetic symmetry of $\psi_{3,i}$ in the crystalline phase of the silica bilayer.}
\label{fig:loop_psi3}
\end{figure*}

Representative snapshots of the spatial distribution of the 3-fold bond-order parameter between silicons are shown in Fig.~\ref{fig:loop_psi3}, from above the crossover temperature $T_x$ down to the lowest equilibrated temperature.
The color-coding is chosen so as to highlight regions with similar phase of the order parameter, 
while the radius of the particles is proportional to the amplitude $|\psi_{3,j}|$.
We clearly observe patches of particles with similar orientational order, which are visible around the crossover temperature $T_x$ and 
whose typical size grows appreciably with decreasing temperature.
At low temperature, the system is populated by numerous crystalline patches of nanometric size and different orientations.

\begin{figure*}[!tb]
\centering
\includegraphics[width=0.95\textwidth]{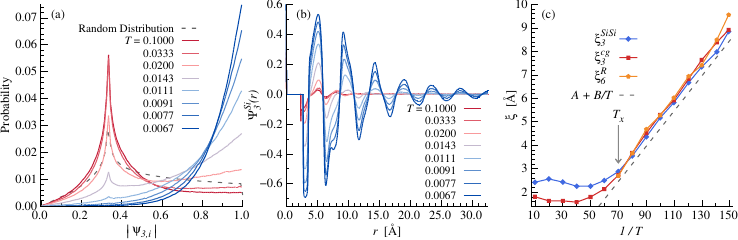}
\caption{
(a) Distribution of the module of the local order parameter for a 3-fold symmetry between \textit{Si} particles at different temperatures. The dashed distribution corresponds to a random distribution of particles.
(b) Spatial correlation function $\Psi_3^{Si}(r)$ of the local order parameter for a 3-fold symmetry between \textit{Si} particles at different temperatures.
(c) Correlation lengths extracted as a function of the inverse temperature: $\xi_3^{SiSi}$ (circles), $\xi_3^{cg}$ (squares), and $\xi_6^R$ (pentagons). 
The spatial correlations and the correlation length are calculated for $N=10^5$ particles.}
\label{fig:psi3}
\end{figure*}

The probability density function $p(|\psi_3|)$ is shown in Fig.~~\ref{fig:psi3}(a).
It shows the progressive crossover from the nearly random arrangements of the closest 3 bonds at high $T$, to orientationally ordered local arrangements at low $T$.
To quantify the spatial extent of the domains, we compute the $n$-fold bond-orientational correlation function
\begin{equation}
  \Psi_n^{Si}(r) = \frac{\langle \sum_{i,j} \, \psi_{n,i} \psi_{n,j}^* \, \delta(r-r_{ij})\rangle}
  {\langle\sum_{i,j} \, \delta(r-r_{ij})\rangle}   \, ,
\end{equation}
where $r_{ij}=\left| \mathbf{r_i} - \mathbf{r_j} \right|$.
We also define in a similar way $\Psi^{cg}_n(r)$ and $\Psi_n^R(r)$ using the respective local order parameters.
We show the temperature variation of $\Psi_3^{Si}(r)$ in Fig.~\ref{fig:psi3}(b).
The correlation function has alternate negative and positive regions, due to the antiferromagnetic-like order between successive shells of neighbors.
Clearly, the absolute value of the correlation increases progressively as $T$ is lowered, confirming the visual impression from Fig.~\ref{fig:loop_psi3}.

To extract the associated correlation length $\xi_n$, we analyze the envelope of $|\Psi_n^{Si}(r)|$ following the procedure described in the Appendix~\ref{sec:appendix:xi}.
We proceed in a similar fashion to compute the correlation length, $\xi^{cg}_3$, of the coarse-grained bond-order parameter and the one, $\xi^{R}_6$, associated to the bond-order between rings.
In Fig.~\ref{fig:psi3}(c), we show the temperature dependence of all these correlation lengths.
We see that 3-fold orientational correlations start growing precisely when $T$ drops below the crossover temperature $T_x$, at which a well-defined network structure forms.
The growth of orientational order is described by a remarkably simple scaling $\xi \sim 1/T$, which accounts, below $T_x$, for the $T$-dependence of all the static correlation lengths.
At the lowest investigated temperature, the correlation length reaches about one nanometer, but even larger correlated domains are observed in Fig.~\ref{fig:loop_psi3}

The above results show that local crystalline order grows steadily in 2$d$ silica upon cooling below the crossover temperature $T_x$.
Interestingly, we do not see any signature of irreversible crystallization (in the sense of quasi-long range order): the crystalline domains break up on the same time scale as the structural relaxation time $\tau$, see Sec.~\ref{sec:liquid:dynamics}, and the system remains an ergodic liquid in the whole temperature range we investigated.
We found that the liquid is not yet thermodynamically metastable with respect to the crystal in the whole temperature range we investigated.
To address this point, we have estimated the melting temperature of the crystal by brute force simulations.
A crystalline sample of $N=2200$ particles was heated up at the target temperature $T$ and then annealed.
During the annealing the amplitude of 3-fold correlations were measured and melting was identified by their drop to a small, finite value.
From these preliminary calculations, it appears that melting always occurs within a time scale accessible to our simulations down to $T=0.0067$, which is the lowest temperature at which we could equilibrate the liquid.
We repeated the analysis for different system sizes at $T=0.0091$ and found no major finite size effects in the melting time.
A more systematic investigation of the equilibrium phase diagram of the model is left to a future study.

\subsection{Mermin-Wagner fluctuations and finite-size effects}
\label{sec:liquid:MW}

We now analyze the dynamics of 2$d$ silica in equilibrium conditions from high temperature down to the glassy regime.
It is well-established that the dynamics of 2$d$ liquids, as measured by standard time-dependent correlation functions, is severely affected by MW fluctuations~\cite{Shiba_Keim_Kawasaki_2018}.
The goal of this section is to account for these effects, thus putting the analysis of the dynamics of the RHH-II model on firm grounds.

To extract meaningful information about the dynamics in the glassy regime, it is possible to remove the effect of MW fluctuations by using cage-relative particle displacements~\cite{Shiba_Yamada_Kawasaki_Kim_2016, Shiba_Keim_Kawasaki_2018}.
In practice, we first define the displacement of the cage of the $i$-th particle between the moments of time $t$ and $t+t_0$,
\begin{equation}
\Delta\mathbf{\tilde{r}}_{i}(t, t_0) = \frac{1}{N_i} \sum_{j=1}^{N_i} \left[ \mathbf{r}_j(t+t_0) - \mathbf{r}_j(t_0) \right] \, ,
\end{equation}
where the sum runs over the $N_i$ neighbors of the $i$-th particle.
The cage-relative displacement $\mathbf{r}_i(t+t_0) - \mathbf{r}_i(\tau) - \Delta\mathbf{\tilde{r}}_{i}(t, t_0)$ is then used to define the corresponding cage-relative correlation functions.

\begin{figure}[!tb]
\centering
\includegraphics[width=0.95\linewidth]{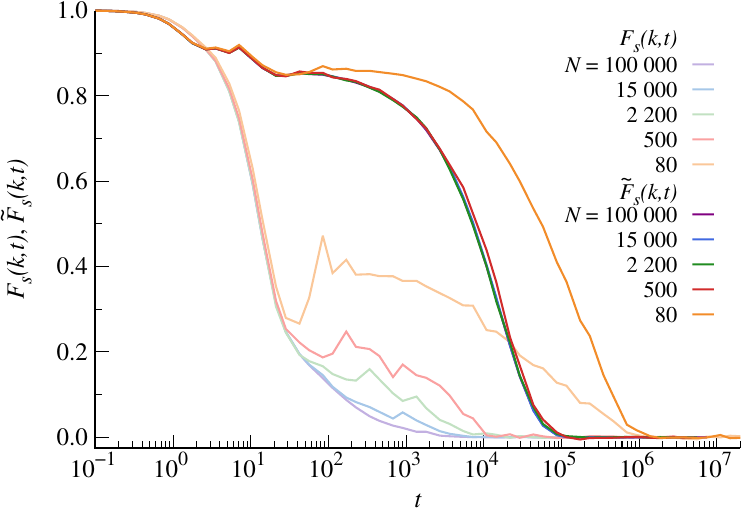}
\caption{Self intermediate scattering function in its standard $F_s(k,t)$ and cage-relative $\tilde{F}_s(k,t)$ version for different system sizes, as indicated in the figure.
The temperature is $T=0.0100$, while the chosen wave-number is $k_2$.}
\label{fig:MW_fskt}
\end{figure}

Typical results of this procedure are illustrated in Fig.~\ref{fig:MW_fskt} from simulations at $T=0.0100$, slightly below $T_x$.
The observable of interest is the self intermediate scattering function, which is calculated both using the bare coordinates of the \textit{Si} particles
\begin{equation}
F_{s}(\mathbf{k},t) = \left\langle \frac{1}{N_{\alpha}} \sum_{j=1}^{N_{\alpha}} e^{i \mathbf{k} \cdot \left[ \mathbf{r}_j(t) - \mathbf{r}_j(0) \right] } \right\rangle  \, ,
\end{equation}
and the cage-relative displacements
\begin{equation}
\tilde{F}_{s}(\mathbf{k},t) = \left\langle \frac{1}{N_{\alpha}} \sum_{j=1}^{N_{\alpha}} e^{ i \mathbf{k} \cdot \left[ \mathbf{r}_j(t) - \mathbf{r}_j(0) - \Delta\mathbf{\tilde{r}}_{j}(t, 0) \right] } \right\rangle  \, .
\end{equation}

As found in close-packed glass-forming liquids~\cite{Flenner_Szamel_2015, Shiba_Yamada_Kawasaki_Kim_2016}, the bare $F_s(k,t)$ displays strong finite size effects and MW fluctuations progressively wash out the glassy dynamics as $N$ increases.
Note that the oscillations of $F_s(k,t)$ at intermediate times are not due to statistical noise: we checked that the time of the first dip in $F_s(k,t)$ displays the $N$ dependence expected from MW fluctuations.
When using cage-relative displacements, instead, finite size effects almost completely disappear and the correlation function reveals a genuine glassy dynamics, with a well-defined relaxation time $\tau_\alpha(k)$ at $\tilde{F}_s(k,\tau_\alpha(k))=1/e$.
We found qualitatively similar results for the mean squared displacement (not shown): finite size effects in the cage-relative mean squared displacement disappear above a few hundreds particles, around $N=200$.
This contrasts with the analysis of Roy \textit{et al.} \cite{Roy_Heuer_2022}, who suggested instead that $N=80$ would be enough to remove major finite size effects, as found in 3$d$ liquid silica~\cite{Saksaengwijit_Heuer_2007}.
We tentatively attribute this discrepancy to the fact that Roy \textit{et al.} mostly focused on structural quantities, on which finite size effects are less visible.

\subsection{Glassy dynamics and dynamic crossover}
\label{sec:liquid:dynamics}

\begin{figure}[!ht]
\includegraphics[width=0.95\linewidth]{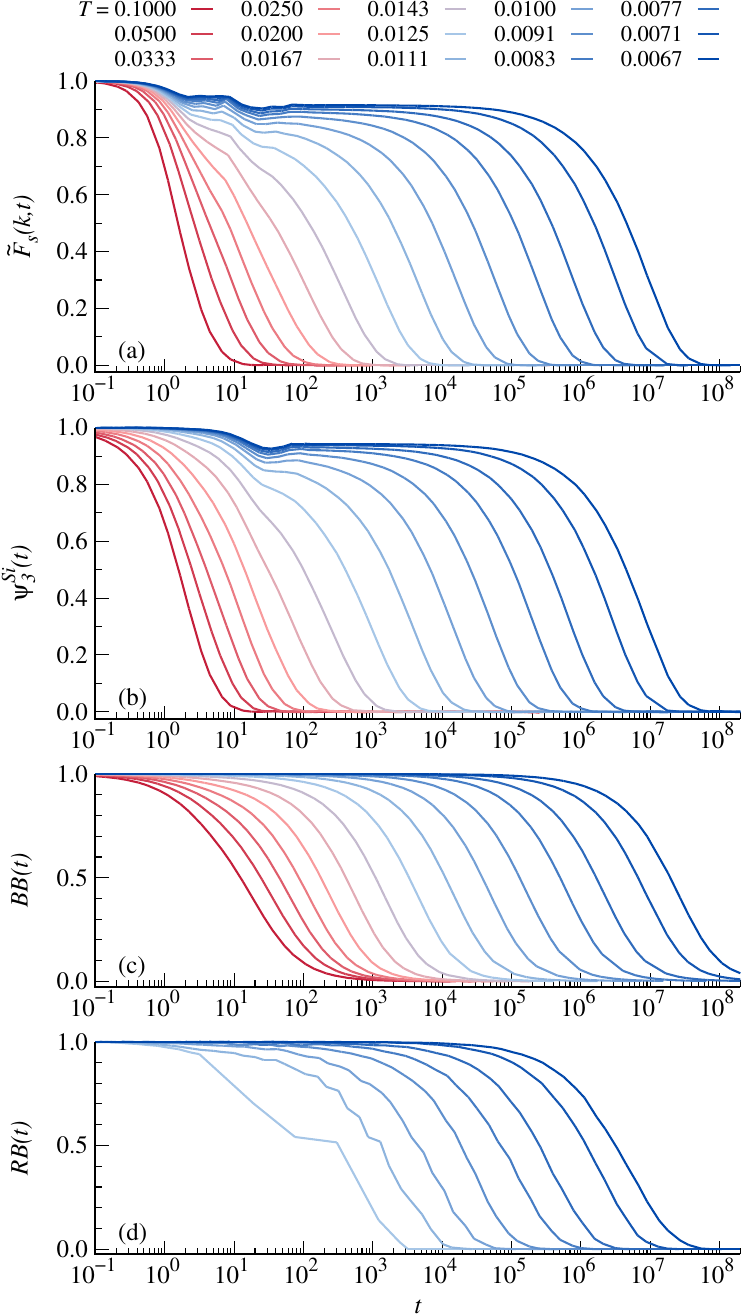}
\caption{(a) Cage-relative self intermediate scattering functions (at $k=k_2$), (b) orientational correlation functions for the 3-fold order parameter between \textit{Si-Si} particles, (c) bond breaking correlation functions and (d) ring breaking correlation functions for different temperatures, as indicated in the figure. 
Note that the ring breaking correlation functions can only be computed for temperatures $T\le0.0125$, due to the lack of a well-defined network structure at higher temperatures. 
The cage-relative self intermediate scattering functions, orientation correlation functions and bond breaking correlation functions are calculated for $N=10^5$ particles, while the ring breaking correlation functions for $N=2200$ particles.}
\label{fig:phis}
\end{figure}


We now turn our attention to the equilibrium dynamics of the RHH-II model, characterizing motion over increasing length scales.
We remove the effect of MW fluctuations either by using cage-relative displacements or considering correlation functions that are independent of MW fluctuations by construction, \textit{e.g.}, bond-orientational and bond-breaking correlation functions.
We report results for the largest available system size, $N=10^5$, and over a considerably larger temperature range than in previous work~\cite{Roy_Heuer_2022}.
Note that, contrary to Roy \textit{et al.}, we will not analyze the diffusion coefficients extracted from the bare MSD, because they are ill-defined in 2$d$~\cite{Shiba_Keim_Kawasaki_2018} and may decouple from structural relaxation times~\cite{li_long-wavelength_2019}.

We start by analyzing the cage-relative self intermediate scattering functions $\tilde{F}_s(k,t)$ evaluated at
$k=k_2$, which corresponds to the main peak of the $S_{SiO}(k)$ structure factor.
The corresponding length scale is around the typical \textit{Si-O} bond distance.
We show these results for the \textit{Si} particles in Fig.~\ref{fig:phis}~(a).
The correlation functions display the typical glassy behavior, with structural relaxation taking on longer time scales as $T$ decreases.
Note the appearance of a plateau, only slightly masked by short-time vibrational motions, around the onset temperature $T_o \approx 0.20 > T_x$.

The same qualitative behavior can be seen in Fig.~\ref{fig:phis}~(b) for the correlation function of the bond-order parameter between silicon particles
$$
\Psi_3(t) = \frac{\langle \sum_{i=1}^N \Psi_{3,i}(t) \Psi_{3,i}^*(0)\rangle}{\langle \sum_{i=1}^N \Psi_{3,i}(0) \Psi_{3,i}^*(0)\rangle} \, .
$$
Interestingly, the slowing down of translational and orientational degrees of freedom proceeds on the same footing across the whole range of temperatures, and occurs on the same time scale at low temperatures.
This behavior is at variance with 2$d$ liquids with close-packed local arrangements, for which translational and orientational degrees of freedom tend to be coupled only at low temperature~\cite{Scalliet_Guiselin_Berthier_2022}.

It is useful to compare these results with correlation functions that measure more directly the dynamics of the bonds.
We focus here on bonds between silicon and oxygen particles and compute the probability that such a bond persists up to time $t$.
The corresponding bond-breaking correlation function is shown in Fig.~\ref{fig:phis}(c).
Although this function does not show any plateau at intermediate times, its slowing down upon cooling is similar to $\tilde{F}_s(k,t)$.

Moving on to larger length scales, we analyze the dynamics of the rings, which are expected to constitute a key structural element of 2$d$ silica networks.
Similarly to the bond-breaking correlation function, we define the ring correlation function as the probability that a ring has not changed its identity, \textit{i.e.}, the ordered sequence of indices of the particles composing the ring, after a time $t$.
As shown in Fig.~\ref{fig:phis}(d), this correlation function is only well-defined below the crossover temperature.
Apart from this, it shows a similar shape as the bond-breaking correlation function, but it relaxes on a shorter time scale.
We found that the two sets of correlation functions can be collapsed onto one another by upscaling the time by about a factor 6 in the ring breaking correlation function (note that 12 bonds are involved in a 6-fold ring).

\begin{figure}[!tb]
\centering
\includegraphics[width=0.8\linewidth]{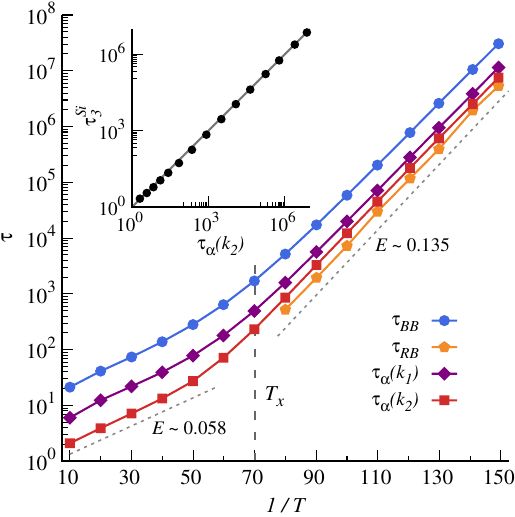}
\caption{Correlation times of the bond breaking correlation function ($\tau_{BB}$, circles), of the ring breaking correlation function ($\tau_{RB}$, pentagons) and cage-relative $\alpha$-relaxation times for silicon atoms, $\tau_\alpha(k_1)$ (diamonds) and $\tau_\alpha(k_2)$ (squares). All quantities are calculated for $N=10^5$ particles, except the correlation time for the ring breaking correlation functions, for which we use $N=2200$ particles. The inset shows the orientational relaxation time $\tau_3$ as a function of $\tau_\alpha(k_2)$.}
\label{fig:taus}
\end{figure}

The structural relaxation times extracted from the above correlation functions are gathered in Fig.~\ref{fig:taus} in an Arrhenius representation.
We also include in the figure the relaxation times obtained from the $\tilde{F}_s(k,t)$ calculated at $k=k_1$, i.e., close to the first sharp diffraction peak of $S_{SiSi}(k)$, as well as the orientational relaxation times from $\Psi_3(t)$, see inset of Fig.~\ref{fig:taus}.
All the relaxation times grow hand in hand with decreasing temperature.
Three regimes are clearly visible from our data.
Both at high and low temperature, the relaxation times are well described by an Arrhenius law, $\tau_0 \exp{E/T}$.
The apparent activation energies are $E\simeq 0.058$ and $E\simeq 0.135$, at high and low temperatures, respectively.
Thus, 2$d$ silica is a perfect example of strong glass-former in the Angell classification, consistent with the findings of Roy \textit{et al.}~\cite{Roy_Heuer_2022}, which were based, however, on the diffusion coefficient and covered a smaller temperature regime.
Between the two Arrhenius regimes, there is a crossover regime centered around $T_x$, over which the liquid behaves as a fragile glass-former.
This regime is, however, very narrow and only covers a decade in relaxation times.
Overall, our results puts the analysis of Roy \textit{et al.} on a firmer basis and clarify the nature of the distinct dynamics regimes in liquid 2$d$ silica.
In particular, while we see a clear strong behavior at low temperature, extended over at least four decades in relaxation times, there is no evidence of two distinct ``fragile'' and ``mixed'' regimes, as discussed in Ref.~\cite{Roy_Heuer_2022}.

\section{Discussion}
\label{sec:discussion}

The results obtained in this work allow us to revisit the physics of network-forming liquids and highlight some differences between the glass transition in two and three dimensions.
We briefly discuss these points below and identify a few directions for future work.

\textit{Locally favored structures and medium range order} -- Triangular arrangements constitute the locally favored structure of 2$d$ silica, while 6-fold rings are the favored motifs on an intermediate length scale.
A striking observation is that the orientational order associated to either kinds of structures becomes increasingly correlated upon cooling below the crossover temperature $T_x$ and the orientational correlation lengths reach about 10~$\AA$ at the lowest temperature where we could reach equilibrium.
The growth of the orientational order may arise from the relatively long-ranged nature of the interactions, which induce rigid structural units propagating order over an intermediate range, similar to what is found in supercooled water~\cite{Shi_Tanaka_2018}. 
The correlated crystalline domains we observe at low temperatures are not compatible with a simple continuous random network picture of the structure, but are instead reminiscent of the old concept of cybotactic grouping developed in the late 1930's, see Ref.~\cite{Wright_2020} for a recent review on the topic.
Although these latter ideas never turned into a quantitative and predictive theoretical description, they seem in better qualitative agreement with the structure of well-annealed 2$d$ silica than those based on continuous random networks.

\textit{Crystalline local order and lack of metastability} --
The presence of local crystalline domains raises questions about the connection with crystallization, in the sense of quasi-long range order.
Our preliminary estimate of the melting temperature is lower than lowest equilibrated temperature, and much lower than $T_x$.
In this regard, 2$d$ amorphous silica behaves similarly to its 3$d$ counterpart, which is already a highly viscous liquid at the melting temperature.
We emphasize, however, that the crystalline domains we observe at low temperature always relax on a time scale comparable to the structural relaxation time, and do not signal incipient crystallization.
We thus observe a striking situation, in which the liquid ``steals'' its local structure from the crystal, doing so well before becoming thermodynamically metastable.
The behavior is thus different from the one of supercooled liquids that instead ``borrow'' their local structure from the complex unit cell of an underlying stable crystal phase~\cite{Pedersen_Douglass_Harrowell_2021}.
A more accurate determination of the crystallization kinetics and of the phase diagram of 2$d$ silica models is needed to corroborate these preliminary results.

\textit{Difference between glassy physics in 2$d$ and 3$d$} --
Recent work has shown that the dynamics of glass-forming liquids appear similar in 2$d$ and 3$d$, once expressed in terms of cage-relative displacements or correlation functions that are insensitive to MW fluctuations~\cite{Tarjus_2017}.
However, several glass-forming liquids in 2$d$, such as bidisperse and polydisperse colloidal particles~\cite{Tanaka_Kawasaki_Shintani_Watanabe_2010, Kawasaki_Tanaka_2011} and models of metallic alloys~\cite{Hu_Tanaka_Wang_2017, Massa_Leporini_Puosi_2019}, display locally ordered domains whose symmetry is \textit{the same} as the stable crystalline phase, while this feature is believed to be mostly absent in good glass-forming liquids in three dimensions.
We observed a similar discrepancy in silica: in 2$d$, the locally favored structures form correlated domains at low temperatures, while there is no evidence of growing orientational order of the tetrahedral units in 3$d$~\cite{Niu_Piaggi_Invernizzi_Parrinello_2018}.
Moreover, subtle forms of compositional order present in 2$d$ polydisperse systems~\cite{Tong_Tanaka_2023} disappear in their 3$d$ counterpart~\cite{Coslovich_Ozawa_Berthier_2018}.
Overall, these findings suggest that while the dynamics of glass-forming liquids appear similar in 2$d$ and 3$d$, their structures may not.

\begin{figure}[!t]
\centering
\includegraphics[width=0.95\linewidth]{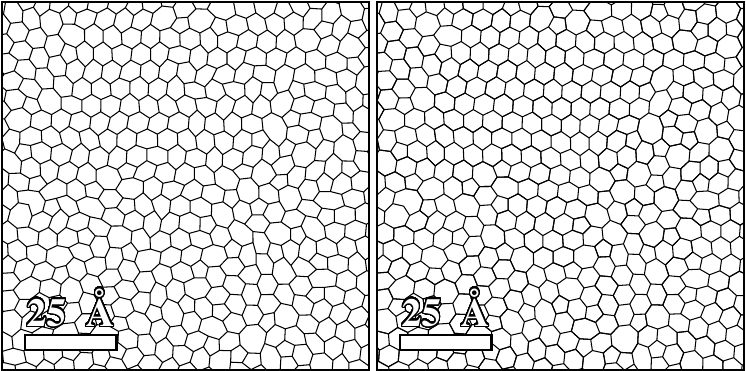}
\caption{Ring structure of a configuration in thermal equilibrium at $T=0.0067$ (left) and its inherent structure (right).}
\label{fig:conc_vs_is}
\end{figure}

\textit{Fragile-to-strong crossover} --
The glassy dynamics of network-forming liquids, such as silica and water, is often interpreted in terms of a fragile-to-strong crossover~\cite{Xu_Kumar_Buldyrev_Chen_Poole_Sciortino_Stanley_2005}: below the onset temperature, the dynamics first slows down rapidly, \textit{e.g.}, following the power law behavior predicted by mode-coupling theory, and then crosses over to an Arrhenius law.
For silica, this picture was first supported by simulation data with the BKS forcefield~\cite{Horbach_Kob_1999}, but the connection with the mode-coupling crossover has been challenged by more recent works~\cite{Coslovich_Ninarello_Berthier_2019, Luo_Robinson_Pihlajamaa_Debets_Royall_Janssen_2022}.
We found that the fragile regime of the RHH-II model is very narrow, covering only about a decade of relaxation times, and no power law can reasonably fit those data.
A recent numerical study of the of the RHH model~\cite{Roy_Heuer_2022} has connected instead the fragile-to-strong crossover to the presence of a low-energy cutoff in the potential energy landscape, suggesting the existence of an amorphous ground state. 
While our results for the inherent structure energies are compatible with those of Ref.~\cite{Roy_Heuer_2022}, our structural analysis indicates that 2$d$ silica accumulates a growing degree of crystallinity, raising doubts about the nature of the putative amorphous ground state at low temperatures.
This is underscored by the snapshot in Fig.~\ref{fig:conc_vs_is} of a representative inherent structure at the lowest investigated temperature, showing crystalline domains at least as wide as 50$\AA$.

\textit{Two-state picture} --
It is tempting to interpret the dynamic crossover observed in 2$d$ silica using the phenomenological two-state picture introduced by Tanaka and collaborators to account for similar observations in 3$d$ silica and supercooled water~\cite{Shi_Tanaka_2018, Shi_Russo_Tanaka_2018}.
In this picture, the system is depicted as a regular mixture of two liquids with distinct local structural states.
The connection with the dynamics is introduced by assuming that each state has its own energy barrier for relaxation, leading to two distinct Arrhenius laws at high and low temperature.
The intermediate fragile regime is then a crossover region, which merely interpolates between the two Arrhenius regimes, while the peak of $C_V$ and the inflection of $e_{IS}(T)$ reflect the coexistence between two distinct structural states.
From a preliminary analysis, which we will develop further in a future study, we found that this interpretation is at least qualitatively compatible with our data.
However, this picture neglects that locally favored structures do not behave as independent structural units: their orientations become increasingly correlated as the temperature drops below $T_x$.

\textit{Elementary rearrangements and dynamic heterogeneity} -- 
Turning our attention to thermodynamic theories of cooperative dynamics~\cite{Ozawa_Scalliet_Ninarello_Berthier_2019}, one may expect that $\log(\tau_\alpha) \sim \xi^\phi / T$, where $\xi$ is a static correlation length measuring the size of the cooperative regions and $\phi$ is a dynamic exponent.
By combining the trends found at low temperatures in Fig.~\ref{fig:psi3}(c) and Fig.~\ref{fig:taus}, we obtain instead $\log(\tau_\alpha) \sim \xi$ for the RHH-II model.
Clearly, the local crystalline domains do not behave as cooperatively rearranging regions as envisioned by thermodynamic theories: their relaxation is likely mediated by numerous rearrangements due to defects in the their surrounding.
The presence of long-lived, favored structures suggests nonetheless that dynamic heterogeneity in 2$d$ silica has a strong structural origin, which could be quantified within the isoconfigurational ensemble approach~\cite{Widmer-Cooper_Harrowell_Fynewever_2004}.
Rationalizing the different relaxation mechanisms at play in glass-forming liquids is a key challenge for theories of the glass transition~\cite{Scalliet_Guiselin_Berthier_2022}, which may find in 2$d$ silica a clean, if somewhat peculiar, test bench.

\section{Conclusions}

In this work, we studied simple but realistic computational models for 2$d$ silica.
Starting from the RHH model~\cite{roy_modelling_2018}, we optimized the interaction parameters to improve the agreement with the in-plane structure of experimental silica bilayers.
The so-obtained RHH-II model better reproduces the ring statistics of experimental silica bilayers, yields dissortative network structures compatible with experiments, and is more efficient to simulate than the original RHH model.
By contrast, a short-range model based on LJ interactions, similar to the one proposed by Coslovich and Pastore for 3$d$ amorphous silica~\cite{coslovich_dynamics_2009}, did not provide a satisfactory structural description.

We then investigated the nature of the glass transition in the RHH-II model by slowly cooling the liquid from high temperature.
From well-equilibrated simulations, we computed a range of structural and dynamic observables, carefully removing from the latter the effect of MW fluctuations~\cite{Shiba_Keim_Kawasaki_2018} to reveal the underlying glassy dynamics.
Our analysis pinpoints a unique crossover temperature $T_x$ that characterizes both the thermodynamic anomalies and the dynamic crossover to a low-temperature Arrhenius regime.
Below $T_x$, the network liquid forms transient crystalline domains whose size increases with decreasing temperature.
A preliminary estimate of the melting temperature indicates, however, that the liquid is not metastable with respect to the crystal in the studied range of temperatures.

Our observations indicate that, despite the lack of strictly long-range order~\cite{Mermin_1968}, there is much more order in well-annealed 2$d$ silica than expected from previous experimental and computational studies~\cite{Sadjadi_Bhattarai_Drabold_Thorpe_Wilson_2017}.
Experimental samples of silica bilayers often show a coexistence of amorphous and crystalline patches.
We suggest that these heterogeneities may be partly due to fluctuations in the vapor deposition rate across the sample, or to inhomogeneities of the substrate, corresponding to a different degree of annealing across the sample.
Generalizing these observations to a broader range of amorphous two-dimensional materials of experimental relevance is a challenge for future computational studies.


\appendix

\setlength{\tabcolsep}{4pt}
\begin{table*}[!t]
\centering
\begin{tabular}{c c c c c}
\toprule
\textbf{Sample} & \textbf{Reference} & \textbf{Type} & \textbf{Substrate} & \textbf{Appearing also as}  \\
\midrule
\midrule
\textbf{A} 	& Tab.~S3 in~\cite{lichtenstein_atomic_2012-1} (SI) 				& Atoms coordinates		& Ru(0001)	& Fig.~2c in~\cite{kumar_ring_2014} \\
\textbf{B}		& Tab.~S4 in~\cite{lichtenstein_atomic_2012-1} (SI)					& Atoms coordinates		& Ru(0001)	& Fig.~2$d$ in~\cite{kumar_ring_2014}		\\
\textbf{C}		& Tab.~S2 in~\cite{lichtenstein_crystalline-vitreous_2012} (SI)		& Atoms coordinates		& Ru(0001)	& Fig.~2b in~\cite{kumar_ring_2014}		\\
\textbf{D}		& Fig.~2a in~\cite{kumar_ring_2014} (SI)		& Ring structures		& Graphene on copper foil		& $-$	\\
\textbf{E}		& Fig.~2e in~\cite{kumar_ring_2014} (SI)		& Ring structures		& Graphene on copper foil		& $-$	\\
\bottomrule
\end{tabular}
\caption{Experimental samples used in this work.}
\label{tab_exp_sample}
\end{table*}

\section{Experimental samples}
\label{sec:appendix:exp:samples}
Although silica bilayers were synthesized as early as in 2012~\cite{huang_direct_2012,lichtenstein_atomic_2012}, only a handful of experimental configurations have been made publicly available.
Moreover, the information is scattered over several papers, which may lead to some uncertainty when computing the statistical properties of amorphous 2$d$ silica.

To bring some order, we present in Table~\ref{tab_exp_sample} the list of the experimental configurations we used to determine the reference experimental data in this paper.
To the best of our knowledge, these are the only amorphous experimental configurations for which the atoms' positions are available or can be easily extracted.
For each configuration, we also report the ring statistics and the assortativity, in order to provide the reader with the complete set of raw data.
Of the five configurations we used, only three of them explicitly report the atoms' coordinates and can be used for the calculation of the RDFs and the BADs.
The other two configurations are available as figures with colored rings.
For these last two configurations, we digitized the positions of each ring vertex, \textit{i.e.}, the positions of the \textit{Si} particles, and we scaled those coordinates so that the average \textit{Si-Si} distance equals the experimental one.
Given the somewhat arbitrary procedure used to obtain these coordinates, we did not use them for the RDFs and BADs calculations.

\begin{table*}[!tb]
\centering
\begin{tabular}{c c c c c c c c c}
\toprule
\multirow{2}{4em}{\textbf{Sample}} & \multicolumn{7}{c}{\textbf{Number of rings with $n$ sides}} & \multirow{2}{4em}{\textbf{Total}} \\
& \textbf{$n=4$} & \textbf{$n=5$} & \textbf{$n=6$} & \textbf{$n=7$} & \textbf{$n=8$} & \textbf{$n=9$} & \textbf{$n=10$} & \\
\midrule
\midrule
\textbf{A} 	& $3$    & $23$    & $39$    & $9$    & $8$   & $0$  & $0$  	& $82$   \\
\textbf{B}		& $5$    & $17$    & $28$    & $13$  & $2$   & $2$  & $0$ 	& $67$   \\
\textbf{C}		& $11$  & $80$    & $130$  & $46$  & $21$  & $2$  & $0$ 	& $290$  \\
\textbf{D}		& $17$  & $121$  & $186$  & $94$  & $22$  & $3$  & $1$ 	& $444$  \\
\textbf{E}		& $15$  & $118$ 	& $204$  & $90$  & $19$  & $2$  & $1$ 	& $449$  \\
\midrule
\multicolumn{2}{l}{\textbf{Ring statistics}} & & & & & & & \\
\textbf{Mean}		& $0.0383$	& $0.2695$ 	& $0.4407$  	& $0.1892$  	& $0.0541$  	& $0.0068$  	& $0.0015$ 	&  \\
\textbf{Std. dev.}		& $0.0043$  	& $0.0034$ 	& $0.0091$  	& $0.0142$  	& $0.0083$  	& $0.0028$  	& $0.0005$ 	&  \\
\bottomrule
\end{tabular}
\caption{Experimental ring statistics for the different experimental samples used in this work.}
\label{tab_exp_ring_statistics}
\end{table*}

In Table~\ref{tab_exp_ring_statistics} and~\ref{tab_exp_assortativity}, we provide the raw data we used for the calculation of the experimental ring statistics and assortativity.
Each quantity is calculated as a weighted average over the configurations, using as weights the total number of rings in each configuration for the average ring statistics and the total number of rings not touching the outer edge for the average assortativity.
We provide both sets of weights in their respective tables.

\begingroup
\setlength{\tabcolsep}{4pt}
\begin{table}[!b]
\centering
\begin{tabular}{c c c}
\toprule
\textbf{Sample} & \textbf{Assortativity} & \textbf{Weight} \\
\midrule
\midrule
\textbf{A} 	& $-0.663$ 	& $74$ \\
\textbf{B}		& $+0.066$ 	& $58$ \\
\textbf{C}		& $-0.256$ 	& $281$ \\
\textbf{D}		& $-0.218$ 	& $434$ \\
\textbf{E}		& $-0.167$ 	& $439$ \\
\midrule
\textbf{Average}	& $-0.2216 \pm 0.0637$ & \\
\bottomrule
\end{tabular}
\caption{Assortativity coefficients for the different experimental samples used in this work.}
\label{tab_exp_assortativity}
\end{table}
\endgroup

\section{Definition of ring}
\label{sec:appendix:rings}

In silica, a ring is typically defined as an ordered set of silicon atoms satisfying appropriate topological rules~\cite{le_roux_ring_2010}.
The oxygens are only used to define the connectivity between the silicon atoms, and play no role in the ring analysis.
Many different definitions of ring have been proposed over the years~\cite{le_roux_ring_2010}.
These typically build on the presence of a shortest path between atoms, such as in Guttman's~\cite{guttman_ring_1990} or shortest path~\cite{king_ring_1967, franzblau_computation_1991} rings, or on the decomposition of larger rings into more fundamental ones, such as strong~\cite{goetzke_properties_1991}, irreducible~\cite{wooten_structure_2002} or primitive~\cite{yuan_efficient_2002} rings.
It should be emphasized that these ring definitions are not equivalent~\cite{le_roux_ring_2010} and the use of different definitions may led to different conclusions on the same set of data~\cite{tirelli_topological_2023}.
The presence of many nonequivalent ring definitions is mainly due to the inherent difficulty of three-dimensional networks and the absence of clear intuitive examples of rings in such systems.
In this work, we propose a new definition of rings that has its roots in graph theory and has a clear intuitive meaning.

Given a system of silicon and oxygen atoms, we can define a graph in which the nodes represent silicon atoms, while the oxygens define the connections between the nodes.
A graph can always be drawn in a planar representation, i.e. without edge-crossing, on a surface of appropriate genus $g$~\cite{trudeau_introduction_1993}.
The nodes and edges in a planar representation naturally divide the surface in different regions called faces~\cite{trudeau_introduction_1993}.
Since the faces are always well defined and are bounded by an ordered set of nodes, we identify the rings in the system with the faces of the underlying graph.

This definition of rings is very intuitive in a two-dimensional network like the one considered here.
At sufficiently low temperatures, the neighbors of each atoms are well defined, and therefore the representation of the system in a rectangular cell with periodic boundary conditions is already a planar representation of the corresponding graph.
A rectangular cell with periodic boundary conditions is isomorphic to the surface of a torus, i.e. a surface of genus $g=1$.
Therefore, the voids we observe in a given configuration of the system are exactly equal to the faces and every set of atoms surrounding one of these regions is a ring according to our definition.
In principle, this definition can also be used for three-dimensional systems, if one of their planar representation can be obtained.
We implemented an algorithm to identify the rings as described above in the library \texttt{l2dr}~\cite{l2dr}, which we used to analyze the network structure in this work.

\section{Calculation of correlation lengths}
\label{sec:appendix:xi}

\begin{figure}[!tb]
\centering
\includegraphics[width=0.95\linewidth]{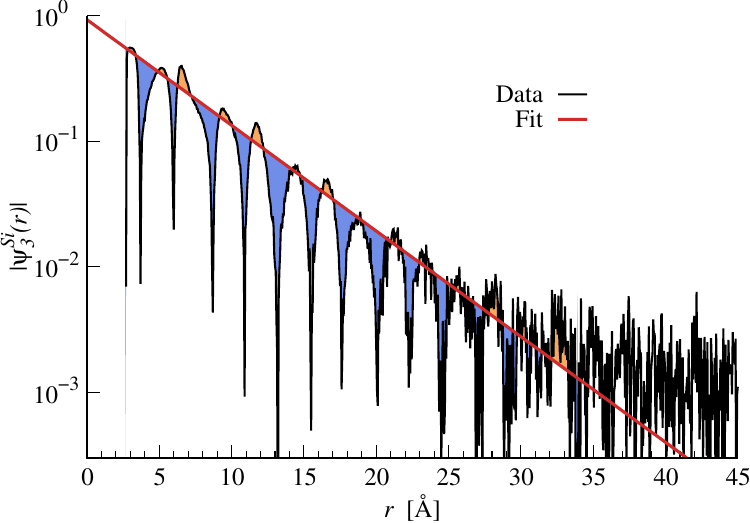}
\caption{Example of the fitting procedure used to extract the correlation length from the curves in~Fig.~\ref{img_rdf} (a). Here $r_{max} = 34$ and the cost function is the sum of twice ($\eta = 2$) the area of the orange regions plus the area of the blue regions.}
\label{fig:psi3_fit}
\end{figure}

To extract the correlation length $\xi_n$ associated to $n$-fold bond-orientational order, we fitted the whole  $|\Psi_n^{Si}(r)|$ to an exponential function $A\exp{-r/\xi_n}$, but giving more weight to the points above the fitting function than those below.
This procedure, which captures the trend of the envelope very well, is illustrated in Fig.~\ref{fig:psi3_fit}, 
More precisely, we minimized the cost function given by
\begin{equation}
\chi^2(A,\xi_n) = \sum_{r \in \left[ r_{min};\, r_{max} \right]} \omega(r)\left( - \frac{r}{\xi_n} + \log A - \log \left|\Psi_n(r)\right|\right)\, ,
\end{equation}
where $\left[ r_{min};\, r_{max} \right]$ is the interval over which the fit is carried out and
\begin{equation}
\omega(r) = \begin{cases}1, & \mbox{if }\ A\,e^{-r/\xi_n} \ge \Psi_n(r) \\ -\eta, & \mbox{otherwise}
\end{cases} \, .
\end{equation}
The minimum distance was fixed to the first non-zero value of $\Psi_n(r)$ for the 3-fold correlations between \textit{Si-Si} and \textit{Si-O} particles and $r_{min}\simeq 4.0$ for the 6-fold correlations between rings, while the maximum one was adjusted depending on temperature so as to remove the noisy part of the correlation function at large $r$.
The weighting function $\omega(r)$ allows one to fine tune the relative weight of points above the fitting line compared with the ones below.
We chose $\eta = 2$, since we found the results to be independent of $\eta$ above this value.
Note that even $\eta=1$ would give the same qualitatively similar trends as a function of $T$.

\section*{Acknowledgments} 

We thank L. Costigliola and T. B. Schrøder for useful discussions.
We acknowledge the CINECA computing center for allocation of computational resources.
We also thank the Glass and Time group of the Roskilde University for a generous allocation of computational resources on their GPU cluster.
We acknowledge support from the Fondazione ICSC “Italian Research Center on High-Performance Computing, Big Data, and Quantum Computing” - Spoke 7, Materials and Molecular Sciences - National Recovery and Resilience Plan (PNRR) - funded by MUR Missione 4 - Componente 2 - Investimento 1.4 - Next Generation EU (NGEU). 
The data that support the findings of this study will be openly available after publication in the Materials Cloud platform.


\begin{thebibliography}{77}%
\makeatletter
\providecommand \@ifxundefined [1]{%
 \@ifx{#1\undefined}
}%
\providecommand \@ifnum [1]{%
 \ifnum #1\expandafter \@firstoftwo
 \else \expandafter \@secondoftwo
 \fi
}%
\providecommand \@ifx [1]{%
 \ifx #1\expandafter \@firstoftwo
 \else \expandafter \@secondoftwo
 \fi
}%
\providecommand \natexlab [1]{#1}%
\providecommand \enquote  [1]{``#1''}%
\providecommand \bibnamefont  [1]{#1}%
\providecommand \bibfnamefont [1]{#1}%
\providecommand \citenamefont [1]{#1}%
\providecommand \href@noop [0]{\@secondoftwo}%
\providecommand \href [0]{\begingroup \@sanitize@url \@href}%
\providecommand \@href[1]{\@@startlink{#1}\@@href}%
\providecommand \@@href[1]{\endgroup#1\@@endlink}%
\providecommand \@sanitize@url [0]{\catcode `\\12\catcode `\$12\catcode
  `\&12\catcode `\#12\catcode `\^12\catcode `\_12\catcode `\%12\relax}%
\providecommand \@@startlink[1]{}%
\providecommand \@@endlink[0]{}%
\providecommand \url  [0]{\begingroup\@sanitize@url \@url }%
\providecommand \@url [1]{\endgroup\@href {#1}{\urlprefix }}%
\providecommand \urlprefix  [0]{URL }%
\providecommand \Eprint [0]{\href }%
\providecommand \doibase [0]{https://doi.org/}%
\providecommand \selectlanguage [0]{\@gobble}%
\providecommand \bibinfo  [0]{\@secondoftwo}%
\providecommand \bibfield  [0]{\@secondoftwo}%
\providecommand \translation [1]{[#1]}%
\providecommand \BibitemOpen [0]{}%
\providecommand \bibitemStop [0]{}%
\providecommand \bibitemNoStop [0]{.\EOS\space}%
\providecommand \EOS [0]{\spacefactor3000\relax}%
\providecommand \BibitemShut  [1]{\csname bibitem#1\endcsname}%
\let\auto@bib@innerbib\@empty
\bibitem [{\citenamefont {Zachariasen}(1932)}]{Zachariasen_1932}%
  \BibitemOpen
  \bibfield  {author} {\bibinfo {author} {\bibfnamefont {W.~H.}\ \bibnamefont
  {Zachariasen}},\ }\href {https://doi.org/10.1021/ja01349a006} {\bibfield
  {journal} {\bibinfo  {journal} {J. Am. Chem. Soc.}\ }\textbf {\bibinfo
  {volume} {54}},\ \bibinfo {pages} {3841} (\bibinfo {year}
  {1932})}\BibitemShut {NoStop}%
\bibitem [{\citenamefont {Phillips}\ and\ \citenamefont
  {Thorpe}(1985)}]{Phillips_Thorpe_1985}%
  \BibitemOpen
  \bibfield  {author} {\bibinfo {author} {\bibfnamefont {J.~C.}\ \bibnamefont
  {Phillips}}\ and\ \bibinfo {author} {\bibfnamefont {M.~F.}\ \bibnamefont
  {Thorpe}},\ }\href {https://doi.org/10.1016/0038-1098(85)90381} {\bibfield
  {journal} {\bibinfo  {journal} {Solid State Commun.}\ }\textbf {\bibinfo
  {volume} {53}},\ \bibinfo {pages} {699} (\bibinfo {year} {1985})}\BibitemShut
  {NoStop}%
\bibitem [{\citenamefont {Gupta}\ and\ \citenamefont
  {Cooper}(1990)}]{Gupta_Cooper_1990}%
  \BibitemOpen
  \bibfield  {author} {\bibinfo {author} {\bibfnamefont {P.~K.}\ \bibnamefont
  {Gupta}}\ and\ \bibinfo {author} {\bibfnamefont {A.~R.}\ \bibnamefont
  {Cooper}},\ }\href {https://doi.org/10.1016/0022-3093(90)90768-H} {\bibfield
  {journal} {\bibinfo  {journal} {J. Non-Cryst. Solids}\ }\textbf {\bibinfo
  {volume} {123}},\ \bibinfo {pages} {14} (\bibinfo {year} {1990})}\BibitemShut
  {NoStop}%
\bibitem [{\citenamefont {Gupta}\ and\ \citenamefont
  {Mauro}(2009)}]{Gupta_Mauro_2009}%
  \BibitemOpen
  \bibfield  {author} {\bibinfo {author} {\bibfnamefont {P.~K.}\ \bibnamefont
  {Gupta}}\ and\ \bibinfo {author} {\bibfnamefont {J.~C.}\ \bibnamefont
  {Mauro}},\ }\href {https://doi.org/10.1063/1.3077168} {\bibfield  {journal}
  {\bibinfo  {journal} {J. Chem. Phys.}\ }\textbf {\bibinfo {volume} {130}},\
  \bibinfo {pages} {094503} (\bibinfo {year} {2009})}\BibitemShut {NoStop}%
\bibitem [{\citenamefont {Horbach}\ and\ \citenamefont
  {Kob}(1999)}]{Horbach_Kob_1999}%
  \BibitemOpen
  \bibfield  {author} {\bibinfo {author} {\bibfnamefont {J.}~\bibnamefont
  {Horbach}}\ and\ \bibinfo {author} {\bibfnamefont {W.}~\bibnamefont {Kob}},\
  }\href {https://doi.org/10.1103/PhysRevB.60.3169} {\bibfield  {journal}
  {\bibinfo  {journal} {Phys. Rev. B}\ }\textbf {\bibinfo {volume} {60}},\
  \bibinfo {pages} {3169} (\bibinfo {year} {1999})}\BibitemShut {NoStop}%
\bibitem [{\citenamefont {Meyer}\ \emph {et~al.}(2004)\citenamefont {Meyer},
  \citenamefont {Horbach}, \citenamefont {Kob}, \citenamefont {Kargl},\ and\
  \citenamefont {Schober}}]{Meyer_Horbach_Kob_Kargl_Schober_2004}%
  \BibitemOpen
  \bibfield  {author} {\bibinfo {author} {\bibfnamefont {A.}~\bibnamefont
  {Meyer}}, \bibinfo {author} {\bibfnamefont {J.}~\bibnamefont {Horbach}},
  \bibinfo {author} {\bibfnamefont {W.}~\bibnamefont {Kob}}, \bibinfo {author}
  {\bibfnamefont {F.}~\bibnamefont {Kargl}},\ and\ \bibinfo {author}
  {\bibfnamefont {H.}~\bibnamefont {Schober}},\ }\href
  {https://doi.org/10.1103/PhysRevLett.93.027801} {\bibfield  {journal}
  {\bibinfo  {journal} {Phys. Rev. Lett.}\ }\textbf {\bibinfo {volume} {93}},\
  \bibinfo {pages} {027801} (\bibinfo {year} {2004})}\BibitemShut {NoStop}%
\bibitem [{\citenamefont {Carré}\ \emph {et~al.}(2007)\citenamefont {Carré},
  \citenamefont {Berthier}, \citenamefont {Horbach}, \citenamefont {Ispas},\
  and\ \citenamefont {Kob}}]{Carre_Berthier_Horbach_Ispas_Kob_2007}%
  \BibitemOpen
  \bibfield  {author} {\bibinfo {author} {\bibfnamefont {A.}~\bibnamefont
  {Carré}}, \bibinfo {author} {\bibfnamefont {L.}~\bibnamefont {Berthier}},
  \bibinfo {author} {\bibfnamefont {J.}~\bibnamefont {Horbach}}, \bibinfo
  {author} {\bibfnamefont {S.}~\bibnamefont {Ispas}},\ and\ \bibinfo {author}
  {\bibfnamefont {W.}~\bibnamefont {Kob}},\ }\href
  {https://doi.org/10.1063/1.2777136} {\bibfield  {journal} {\bibinfo
  {journal} {J. Chem. Phys.}\ }\textbf {\bibinfo {volume} {127}},\ \bibinfo
  {pages} {114512} (\bibinfo {year} {2007})}\BibitemShut {NoStop}%
\bibitem [{\citenamefont {Liu}\ \emph {et~al.}(2019)\citenamefont {Liu},
  \citenamefont {Fu}, \citenamefont {Li}, \citenamefont {Sabri},\ and\
  \citenamefont {Bauchy}}]{Liu_Fu_Li_Sabri_Bauchy_2019}%
  \BibitemOpen
  \bibfield  {author} {\bibinfo {author} {\bibfnamefont {H.}~\bibnamefont
  {Liu}}, \bibinfo {author} {\bibfnamefont {Z.}~\bibnamefont {Fu}}, \bibinfo
  {author} {\bibfnamefont {Y.}~\bibnamefont {Li}}, \bibinfo {author}
  {\bibfnamefont {N.~F.~A.}\ \bibnamefont {Sabri}},\ and\ \bibinfo {author}
  {\bibfnamefont {M.}~\bibnamefont {Bauchy}},\ }\href
  {https://doi.org/10.1557/mrc.2019.47} {\bibfield  {journal} {\bibinfo
  {journal} {MRS Communications}\ }\textbf {\bibinfo {volume} {9}},\ \bibinfo
  {pages} {593} (\bibinfo {year} {2019})}\BibitemShut {NoStop}%
\bibitem [{\citenamefont {Lichtenstein}\ \emph
  {et~al.}(2012{\natexlab{a}})\citenamefont {Lichtenstein}, \citenamefont
  {Büchner}, \citenamefont {Yang}, \citenamefont {Shaikhutdinov},
  \citenamefont {Heyde}, \citenamefont {Sierka}, \citenamefont {Włodarczyk},
  \citenamefont {Sauer},\ and\ \citenamefont
  {Freund}}]{lichtenstein_atomic_2012}%
  \BibitemOpen
  \bibfield  {author} {\bibinfo {author} {\bibfnamefont {L.}~\bibnamefont
  {Lichtenstein}}, \bibinfo {author} {\bibfnamefont {C.}~\bibnamefont
  {Büchner}}, \bibinfo {author} {\bibfnamefont {B.}~\bibnamefont {Yang}},
  \bibinfo {author} {\bibfnamefont {S.}~\bibnamefont {Shaikhutdinov}}, \bibinfo
  {author} {\bibfnamefont {M.}~\bibnamefont {Heyde}}, \bibinfo {author}
  {\bibfnamefont {M.}~\bibnamefont {Sierka}}, \bibinfo {author} {\bibfnamefont
  {R.}~\bibnamefont {Włodarczyk}}, \bibinfo {author} {\bibfnamefont
  {J.}~\bibnamefont {Sauer}},\ and\ \bibinfo {author} {\bibfnamefont {H.-J.}\
  \bibnamefont {Freund}},\ }\href {https://doi.org/10.1002/anie.201107097}
  {\bibfield  {journal} {\bibinfo  {journal} {Angew. Chem. Int. Ed.}\ }\textbf
  {\bibinfo {volume} {51}},\ \bibinfo {pages} {404} (\bibinfo {year}
  {2012}{\natexlab{a}})}\BibitemShut {NoStop}%
\bibitem [{\citenamefont {Huang}\ \emph {et~al.}(2012)\citenamefont {Huang},
  \citenamefont {Kurasch}, \citenamefont {Srivastava}, \citenamefont
  {Skakalova}, \citenamefont {Kotakoski}, \citenamefont {Krasheninnikov},
  \citenamefont {Hovden}, \citenamefont {Mao}, \citenamefont {Meyer},
  \citenamefont {Smet}, \citenamefont {Muller},\ and\ \citenamefont
  {Kaiser}}]{huang_direct_2012}%
  \BibitemOpen
  \bibfield  {author} {\bibinfo {author} {\bibfnamefont {P.~Y.}\ \bibnamefont
  {Huang}}, \bibinfo {author} {\bibfnamefont {S.}~\bibnamefont {Kurasch}},
  \bibinfo {author} {\bibfnamefont {A.}~\bibnamefont {Srivastava}}, \bibinfo
  {author} {\bibfnamefont {V.}~\bibnamefont {Skakalova}}, \bibinfo {author}
  {\bibfnamefont {J.}~\bibnamefont {Kotakoski}}, \bibinfo {author}
  {\bibfnamefont {A.~V.}\ \bibnamefont {Krasheninnikov}}, \bibinfo {author}
  {\bibfnamefont {R.}~\bibnamefont {Hovden}}, \bibinfo {author} {\bibfnamefont
  {Q.}~\bibnamefont {Mao}}, \bibinfo {author} {\bibfnamefont {J.~C.}\
  \bibnamefont {Meyer}}, \bibinfo {author} {\bibfnamefont {J.}~\bibnamefont
  {Smet}}, \bibinfo {author} {\bibfnamefont {D.~A.}\ \bibnamefont {Muller}},\
  and\ \bibinfo {author} {\bibfnamefont {U.}~\bibnamefont {Kaiser}},\ }\href
  {https://doi.org/10.1021/nl204423x} {\bibfield  {journal} {\bibinfo
  {journal} {Nano Lett.}\ }\textbf {\bibinfo {volume} {12}},\ \bibinfo {pages}
  {1081} (\bibinfo {year} {2012})}\BibitemShut {NoStop}%
\bibitem [{\citenamefont {Yu}\ \emph {et~al.}(2012)\citenamefont {Yu},
  \citenamefont {Yang}, \citenamefont {Anibal~Boscoboinik}, \citenamefont
  {Shaikhutdinov},\ and\ \citenamefont {Freund}}]{yu_support_2012}%
  \BibitemOpen
  \bibfield  {author} {\bibinfo {author} {\bibfnamefont {X.}~\bibnamefont
  {Yu}}, \bibinfo {author} {\bibfnamefont {B.}~\bibnamefont {Yang}}, \bibinfo
  {author} {\bibfnamefont {J.}~\bibnamefont {Anibal~Boscoboinik}}, \bibinfo
  {author} {\bibfnamefont {S.}~\bibnamefont {Shaikhutdinov}},\ and\ \bibinfo
  {author} {\bibfnamefont {H.-J.}\ \bibnamefont {Freund}},\ }\href
  {https://doi.org/10.1063/1.3703609} {\bibfield  {journal} {\bibinfo
  {journal} {Appl. Phys. Lett.}\ }\textbf {\bibinfo {volume} {100}},\ \bibinfo
  {pages} {151608} (\bibinfo {year} {2012})}\BibitemShut {NoStop}%
\bibitem [{\citenamefont {Büchner}\ and\ \citenamefont
  {Heyde}(2017)}]{buchner_two-dimensional_2017}%
  \BibitemOpen
  \bibfield  {author} {\bibinfo {author} {\bibfnamefont {C.}~\bibnamefont
  {Büchner}}\ and\ \bibinfo {author} {\bibfnamefont {M.}~\bibnamefont
  {Heyde}},\ }\href {https://doi.org/10.1016/j.progsurf.2017.09.001} {\bibfield
   {journal} {\bibinfo  {journal} {Prog. Surf. Sci.}\ }\textbf {\bibinfo
  {volume} {92}},\ \bibinfo {pages} {341} (\bibinfo {year} {2017})}\BibitemShut
  {NoStop}%
\bibitem [{\citenamefont {Ormrod~Morley}\ \emph {et~al.}(2020)\citenamefont
  {Ormrod~Morley}, \citenamefont {Thorneywork}, \citenamefont {Dullens},\ and\
  \citenamefont {Wilson}}]{Ormrod_Morley_Thorneywork_Dullens_Wilson_2020}%
  \BibitemOpen
  \bibfield  {author} {\bibinfo {author} {\bibfnamefont {D.}~\bibnamefont
  {Ormrod~Morley}}, \bibinfo {author} {\bibfnamefont {A.~L.}\ \bibnamefont
  {Thorneywork}}, \bibinfo {author} {\bibfnamefont {R.~P.~A.}\ \bibnamefont
  {Dullens}},\ and\ \bibinfo {author} {\bibfnamefont {M.}~\bibnamefont
  {Wilson}},\ }\href {https://doi.org/10.1103/PhysRevE.101.042309} {\bibfield
  {journal} {\bibinfo  {journal} {Phys. Rev. E}\ }\textbf {\bibinfo {volume}
  {101}},\ \bibinfo {pages} {042309} (\bibinfo {year} {2020})}\BibitemShut
  {NoStop}%
\bibitem [{\citenamefont {Wilson}\ \emph {et~al.}(2013)\citenamefont {Wilson},
  \citenamefont {Kumar}, \citenamefont {Sherrington},\ and\ \citenamefont
  {Thorpe}}]{Wilson_Modeling_2013}%
  \BibitemOpen
  \bibfield  {author} {\bibinfo {author} {\bibfnamefont {M.}~\bibnamefont
  {Wilson}}, \bibinfo {author} {\bibfnamefont {A.}~\bibnamefont {Kumar}},
  \bibinfo {author} {\bibfnamefont {D.}~\bibnamefont {Sherrington}},\ and\
  \bibinfo {author} {\bibfnamefont {M.~F.}\ \bibnamefont {Thorpe}},\ }\href
  {https://doi.org/10.1103/PhysRevB.87.214108} {\bibfield  {journal} {\bibinfo
  {journal} {Phys. Rev. B}\ }\textbf {\bibinfo {volume} {87}},\ \bibinfo
  {pages} {214108} (\bibinfo {year} {2013})}\BibitemShut {NoStop}%
\bibitem [{\citenamefont {Roy}\ \emph {et~al.}(2018)\citenamefont {Roy},
  \citenamefont {Heyde},\ and\ \citenamefont {Heuer}}]{roy_modelling_2018}%
  \BibitemOpen
  \bibfield  {author} {\bibinfo {author} {\bibfnamefont {P.~K.}\ \bibnamefont
  {Roy}}, \bibinfo {author} {\bibfnamefont {M.}~\bibnamefont {Heyde}},\ and\
  \bibinfo {author} {\bibfnamefont {A.}~\bibnamefont {Heuer}},\ }\href
  {https://doi.org/10.1039/C8CP01313F} {\bibfield  {journal} {\bibinfo
  {journal} {Phys. Chem. Chem. Phys.}\ }\textbf {\bibinfo {volume} {20}},\
  \bibinfo {pages} {14725} (\bibinfo {year} {2018})}\BibitemShut {NoStop}%
\bibitem [{\citenamefont {Perera}\ and\ \citenamefont
  {Harrowell}(1999)}]{Perera_Harrowell_1999}%
  \BibitemOpen
  \bibfield  {author} {\bibinfo {author} {\bibfnamefont {D.~N.}\ \bibnamefont
  {Perera}}\ and\ \bibinfo {author} {\bibfnamefont {P.}~\bibnamefont
  {Harrowell}},\ }\href {https://doi.org/10.1103/PhysRevE.59.5721} {\bibfield
  {journal} {\bibinfo  {journal} {Phys. Rev. E}\ }\textbf {\bibinfo {volume}
  {59}},\ \bibinfo {pages} {5721} (\bibinfo {year} {1999})}\BibitemShut
  {NoStop}%
\bibitem [{\citenamefont {Shintani}\ and\ \citenamefont
  {Tanaka}(2006)}]{Shintani_Tanaka_2006}%
  \BibitemOpen
  \bibfield  {author} {\bibinfo {author} {\bibfnamefont {H.}~\bibnamefont
  {Shintani}}\ and\ \bibinfo {author} {\bibfnamefont {H.}~\bibnamefont
  {Tanaka}},\ }\href {https://doi.org/10.1038/nphys235} {\bibfield  {journal}
  {\bibinfo  {journal} {Nat. Phys.}\ }\textbf {\bibinfo {volume} {2}},\
  \bibinfo {pages} {200} (\bibinfo {year} {2006})}\BibitemShut {NoStop}%
\bibitem [{\citenamefont {Kawasaki}\ and\ \citenamefont
  {Tanaka}(2011)}]{Kawasaki_Tanaka_2011}%
  \BibitemOpen
  \bibfield  {author} {\bibinfo {author} {\bibfnamefont {T.}~\bibnamefont
  {Kawasaki}}\ and\ \bibinfo {author} {\bibfnamefont {H.}~\bibnamefont
  {Tanaka}},\ }\href {https://doi.org/10.1088/0953-8984/23/19/194121}
  {\bibfield  {journal} {\bibinfo  {journal} {J. Phys.: Condens. Matt.}\
  }\textbf {\bibinfo {volume} {23}},\ \bibinfo {pages} {194121} (\bibinfo
  {year} {2011})}\BibitemShut {NoStop}%
\bibitem [{\citenamefont {Flenner}\ and\ \citenamefont
  {Szamel}(2015)}]{Flenner_Szamel_2015}%
  \BibitemOpen
  \bibfield  {author} {\bibinfo {author} {\bibfnamefont {E.}~\bibnamefont
  {Flenner}}\ and\ \bibinfo {author} {\bibfnamefont {G.}~\bibnamefont
  {Szamel}},\ }\href {https://doi.org/10.1038/ncomms8392} {\bibfield  {journal}
  {\bibinfo  {journal} {Nat. Commun.}\ }\textbf {\bibinfo {volume} {6}},\
  \bibinfo {pages} {7392} (\bibinfo {year} {2015})}\BibitemShut {NoStop}%
\bibitem [{\citenamefont {Shiba}\ \emph {et~al.}(2016)\citenamefont {Shiba},
  \citenamefont {Yamada}, \citenamefont {Kawasaki},\ and\ \citenamefont
  {Kim}}]{Shiba_Yamada_Kawasaki_Kim_2016}%
  \BibitemOpen
  \bibfield  {author} {\bibinfo {author} {\bibfnamefont {H.}~\bibnamefont
  {Shiba}}, \bibinfo {author} {\bibfnamefont {Y.}~\bibnamefont {Yamada}},
  \bibinfo {author} {\bibfnamefont {T.}~\bibnamefont {Kawasaki}},\ and\
  \bibinfo {author} {\bibfnamefont {K.}~\bibnamefont {Kim}},\ }\href
  {https://doi.org/10.1103/PhysRevLett.117.245701} {\bibfield  {journal}
  {\bibinfo  {journal} {Phys. Rev. Lett.}\ }\textbf {\bibinfo {volume} {117}},\
  \bibinfo {pages} {245701} (\bibinfo {year} {2016})}\BibitemShut {NoStop}%
\bibitem [{\citenamefont {Shiba}\ \emph {et~al.}(2018)\citenamefont {Shiba},
  \citenamefont {Keim},\ and\ \citenamefont
  {Kawasaki}}]{Shiba_Keim_Kawasaki_2018}%
  \BibitemOpen
  \bibfield  {author} {\bibinfo {author} {\bibfnamefont {H.}~\bibnamefont
  {Shiba}}, \bibinfo {author} {\bibfnamefont {P.}~\bibnamefont {Keim}},\ and\
  \bibinfo {author} {\bibfnamefont {T.}~\bibnamefont {Kawasaki}},\ }\href
  {https://doi.org/10.1088/1361-648X/aaa8b8} {\bibfield  {journal} {\bibinfo
  {journal} {J. Phys.: Condens. Matt.}\ }\textbf {\bibinfo {volume} {30}},\
  \bibinfo {pages} {094004} (\bibinfo {year} {2018})}\BibitemShut {NoStop}%
\bibitem [{\citenamefont {Roy}\ and\ \citenamefont
  {Heuer}(2022)}]{Roy_Heuer_2022}%
  \BibitemOpen
  \bibfield  {author} {\bibinfo {author} {\bibfnamefont {P.~K.}\ \bibnamefont
  {Roy}}\ and\ \bibinfo {author} {\bibfnamefont {A.}~\bibnamefont {Heuer}},\
  }\href {https://doi.org/10.1063/5.0118797} {\bibfield  {journal} {\bibinfo
  {journal} {J. Chem. Phys.}\ }\textbf {\bibinfo {volume} {157}},\ \bibinfo
  {pages} {174506} (\bibinfo {year} {2022})}\BibitemShut {NoStop}%
\bibitem [{\citenamefont {Saksaengwijit}\ \emph {et~al.}(2004)\citenamefont
  {Saksaengwijit}, \citenamefont {Reinisch},\ and\ \citenamefont
  {Heuer}}]{Saksaengwijit_Reinisch_Heuer_2004}%
  \BibitemOpen
  \bibfield  {author} {\bibinfo {author} {\bibfnamefont {A.}~\bibnamefont
  {Saksaengwijit}}, \bibinfo {author} {\bibfnamefont {J.}~\bibnamefont
  {Reinisch}},\ and\ \bibinfo {author} {\bibfnamefont {A.}~\bibnamefont
  {Heuer}},\ }\href {https://doi.org/10.1103/PhysRevLett.93.235701} {\bibfield
  {journal} {\bibinfo  {journal} {Phys. Rev. Lett.}\ }\textbf {\bibinfo
  {volume} {93}},\ \bibinfo {pages} {235701} (\bibinfo {year}
  {2004})}\BibitemShut {NoStop}%
\bibitem [{\citenamefont {Tarjus}(2017)}]{Tarjus_2017}%
  \BibitemOpen
  \bibfield  {author} {\bibinfo {author} {\bibfnamefont {G.}~\bibnamefont
  {Tarjus}},\ }\href {https://doi.org/10.1073/pnas.1700193114} {\bibfield
  {journal} {\bibinfo  {journal} {Proc. Natl. Acad. Sci.}\ }\textbf {\bibinfo
  {volume} {114}},\ \bibinfo {pages} {2440} (\bibinfo {year}
  {2017})}\BibitemShut {NoStop}%
\bibitem [{\citenamefont {Tanaka}\ \emph {et~al.}(2010)\citenamefont {Tanaka},
  \citenamefont {Kawasaki}, \citenamefont {Shintani},\ and\ \citenamefont
  {Watanabe}}]{Tanaka_Kawasaki_Shintani_Watanabe_2010}%
  \BibitemOpen
  \bibfield  {author} {\bibinfo {author} {\bibfnamefont {H.}~\bibnamefont
  {Tanaka}}, \bibinfo {author} {\bibfnamefont {T.}~\bibnamefont {Kawasaki}},
  \bibinfo {author} {\bibfnamefont {H.}~\bibnamefont {Shintani}},\ and\
  \bibinfo {author} {\bibfnamefont {K.}~\bibnamefont {Watanabe}},\ }\href
  {https://doi.org/10.1038/nmat2634} {\bibfield  {journal} {\bibinfo  {journal}
  {Nat. Mater.}\ }\textbf {\bibinfo {volume} {9}},\ \bibinfo {pages} {324}
  (\bibinfo {year} {2010})}\BibitemShut {NoStop}%
\bibitem [{\citenamefont {Tong}\ and\ \citenamefont
  {Tanaka}(2018)}]{Tong_Tanaka_2018}%
  \BibitemOpen
  \bibfield  {author} {\bibinfo {author} {\bibfnamefont {H.}~\bibnamefont
  {Tong}}\ and\ \bibinfo {author} {\bibfnamefont {H.}~\bibnamefont {Tanaka}},\
  }\href {https://doi.org/10.1103/PhysRevX.8.011041} {\bibfield  {journal}
  {\bibinfo  {journal} {Phys. Rev. X}\ }\textbf {\bibinfo {volume} {8}},\
  \bibinfo {pages} {011041} (\bibinfo {year} {2018})}\BibitemShut {NoStop}%
\bibitem [{\citenamefont {Royall}\ and\ \citenamefont
  {Williams}(2015)}]{Royall_Williams_2015}%
  \BibitemOpen
  \bibfield  {author} {\bibinfo {author} {\bibfnamefont {C.~P.}\ \bibnamefont
  {Royall}}\ and\ \bibinfo {author} {\bibfnamefont {S.~R.}\ \bibnamefont
  {Williams}},\ }\href {https://doi.org/10.1016/j.physrep.2014.11.004}
  {\bibfield  {journal} {\bibinfo  {journal} {Phys. Rep.}\ }\textbf {\bibinfo
  {volume} {560}},\ \bibinfo {pages} {1} (\bibinfo {year} {2015})}\BibitemShut
  {NoStop}%
\bibitem [{\citenamefont {Crowther}\ \emph {et~al.}(2015)\citenamefont
  {Crowther}, \citenamefont {Turci},\ and\ \citenamefont
  {Royall}}]{Crowther_Turci_Royall_2015}%
  \BibitemOpen
  \bibfield  {author} {\bibinfo {author} {\bibfnamefont {P.}~\bibnamefont
  {Crowther}}, \bibinfo {author} {\bibfnamefont {F.}~\bibnamefont {Turci}},\
  and\ \bibinfo {author} {\bibfnamefont {C.~P.}\ \bibnamefont {Royall}},\
  }\href {https://doi.org/10.1063/1.4927302} {\bibfield  {journal} {\bibinfo
  {journal} {J. Chem. Phys.}\ }\textbf {\bibinfo {volume} {143}},\ \bibinfo
  {pages} {044503} (\bibinfo {year} {2015})}\BibitemShut {NoStop}%
\bibitem [{\citenamefont {Zhong}\ and\ \citenamefont
  {Freund}(2022)}]{zhong_two-dimensional_2022}%
  \BibitemOpen
  \bibfield  {author} {\bibinfo {author} {\bibfnamefont {J.-Q.}\ \bibnamefont
  {Zhong}}\ and\ \bibinfo {author} {\bibfnamefont {H.-J.}\ \bibnamefont
  {Freund}},\ }\href {https://doi.org/10.1021/acs.chemrev.1c00995} {\bibfield
  {journal} {\bibinfo  {journal} {Chem. Rev.}\ }\textbf {\bibinfo {volume}
  {122}},\ \bibinfo {pages} {11172} (\bibinfo {year} {2022})}\BibitemShut
  {NoStop}%
\bibitem [{\citenamefont {Coslovich}\ and\ \citenamefont
  {Pastore}(2009)}]{coslovich_dynamics_2009}%
  \BibitemOpen
  \bibfield  {author} {\bibinfo {author} {\bibfnamefont {D.}~\bibnamefont
  {Coslovich}}\ and\ \bibinfo {author} {\bibfnamefont {G.}~\bibnamefont
  {Pastore}},\ }\href {https://doi.org/10.1088/0953-8984/21/28/285107}
  {\bibfield  {journal} {\bibinfo  {journal} {J. Phys.: Condens. Matter}\
  }\textbf {\bibinfo {volume} {21}},\ \bibinfo {pages} {285107} (\bibinfo
  {year} {2009})}\BibitemShut {NoStop}%
\bibitem [{\citenamefont {Bailey}\ \emph {et~al.}(2017)\citenamefont {Bailey},
  \citenamefont {Ingebrigtsen}, \citenamefont {Hansen}, \citenamefont
  {Veldhorst}, \citenamefont {B{\o}hling}, \citenamefont {Lemarchand},
  \citenamefont {Olsen}, \citenamefont {Bacher}, \citenamefont {Costigliola},
  \citenamefont {Pedersen}, \citenamefont {Larsen}, \citenamefont {Dyre},\ and\
  \citenamefont {Schr{\o}der}}]{bailey_rumd_2017}%
  \BibitemOpen
  \bibfield  {author} {\bibinfo {author} {\bibfnamefont {N.}~\bibnamefont
  {Bailey}}, \bibinfo {author} {\bibfnamefont {T.}~\bibnamefont
  {Ingebrigtsen}}, \bibinfo {author} {\bibfnamefont {J.~S.}\ \bibnamefont
  {Hansen}}, \bibinfo {author} {\bibfnamefont {A.}~\bibnamefont {Veldhorst}},
  \bibinfo {author} {\bibfnamefont {L.}~\bibnamefont {B{\o}hling}}, \bibinfo
  {author} {\bibfnamefont {C.}~\bibnamefont {Lemarchand}}, \bibinfo {author}
  {\bibfnamefont {A.}~\bibnamefont {Olsen}}, \bibinfo {author} {\bibfnamefont
  {A.}~\bibnamefont {Bacher}}, \bibinfo {author} {\bibfnamefont
  {L.}~\bibnamefont {Costigliola}}, \bibinfo {author} {\bibfnamefont
  {U.}~\bibnamefont {Pedersen}}, \bibinfo {author} {\bibfnamefont
  {H.}~\bibnamefont {Larsen}}, \bibinfo {author} {\bibfnamefont
  {J.}~\bibnamefont {Dyre}},\ and\ \bibinfo {author} {\bibfnamefont
  {T.}~\bibnamefont {Schr{\o}der}},\ }\href
  {https://doi.org/10.21468/SciPostPhys.3.6.038} {\bibfield  {journal}
  {\bibinfo  {journal} {SciPost Phys.}\ }\textbf {\bibinfo {volume} {3}},\
  \bibinfo {pages} {038} (\bibinfo {year} {2017})}\BibitemShut {NoStop}%
\bibitem [{\citenamefont {Swallen}\ \emph {et~al.}(2007)\citenamefont
  {Swallen}, \citenamefont {Kearns}, \citenamefont {Mapes}, \citenamefont
  {Kim}, \citenamefont {McMahon}, \citenamefont {Ediger}, \citenamefont {Wu},
  \citenamefont {Yu},\ and\ \citenamefont
  {Satija}}]{Swallen_Kearns_Mapes_Kim_McMahon_Ediger_Wu_Yu_Satija_2007}%
  \BibitemOpen
  \bibfield  {author} {\bibinfo {author} {\bibfnamefont {S.~F.}\ \bibnamefont
  {Swallen}}, \bibinfo {author} {\bibfnamefont {K.~L.}\ \bibnamefont {Kearns}},
  \bibinfo {author} {\bibfnamefont {M.~K.}\ \bibnamefont {Mapes}}, \bibinfo
  {author} {\bibfnamefont {Y.~S.}\ \bibnamefont {Kim}}, \bibinfo {author}
  {\bibfnamefont {R.~J.}\ \bibnamefont {McMahon}}, \bibinfo {author}
  {\bibfnamefont {M.~D.}\ \bibnamefont {Ediger}}, \bibinfo {author}
  {\bibfnamefont {T.}~\bibnamefont {Wu}}, \bibinfo {author} {\bibfnamefont
  {L.}~\bibnamefont {Yu}},\ and\ \bibinfo {author} {\bibfnamefont
  {S.}~\bibnamefont {Satija}},\ }\href
  {https://doi.org/10.1126/science.1135795} {\bibfield  {journal} {\bibinfo
  {journal} {Science}\ }\textbf {\bibinfo {volume} {315}},\ \bibinfo {pages}
  {353–356} (\bibinfo {year} {2007})}\BibitemShut {NoStop}%
\bibitem [{\citenamefont {Lichtenstein}\ \emph
  {et~al.}(2012{\natexlab{b}})\citenamefont {Lichtenstein}, \citenamefont
  {Heyde},\ and\ \citenamefont {Freund}}]{lichtenstein_atomic_2012-1}%
  \BibitemOpen
  \bibfield  {author} {\bibinfo {author} {\bibfnamefont {L.}~\bibnamefont
  {Lichtenstein}}, \bibinfo {author} {\bibfnamefont {M.}~\bibnamefont
  {Heyde}},\ and\ \bibinfo {author} {\bibfnamefont {H.-J.}\ \bibnamefont
  {Freund}},\ }\href {https://doi.org/10.1021/jp3062866} {\bibfield  {journal}
  {\bibinfo  {journal} {J. Phys. Chem. C}\ }\textbf {\bibinfo {volume} {116}},\
  \bibinfo {pages} {20426} (\bibinfo {year} {2012}{\natexlab{b}})}\BibitemShut
  {NoStop}%
\bibitem [{\citenamefont {Lichtenstein}\ \emph
  {et~al.}(2012{\natexlab{c}})\citenamefont {Lichtenstein}, \citenamefont
  {Heyde},\ and\ \citenamefont
  {Freund}}]{lichtenstein_crystalline-vitreous_2012}%
  \BibitemOpen
  \bibfield  {author} {\bibinfo {author} {\bibfnamefont {L.}~\bibnamefont
  {Lichtenstein}}, \bibinfo {author} {\bibfnamefont {M.}~\bibnamefont
  {Heyde}},\ and\ \bibinfo {author} {\bibfnamefont {H.-J.}\ \bibnamefont
  {Freund}},\ }\href {https://doi.org/10.1103/PhysRevLett.109.106101}
  {\bibfield  {journal} {\bibinfo  {journal} {Phys. Rev. Lett.}\ }\textbf
  {\bibinfo {volume} {109}},\ \bibinfo {pages} {106101} (\bibinfo {year}
  {2012}{\natexlab{c}})}\BibitemShut {NoStop}%
\bibitem [{\citenamefont {Kumar}\ \emph {et~al.}(2014)\citenamefont {Kumar},
  \citenamefont {Sherrington}, \citenamefont {Wilson},\ and\ \citenamefont
  {Thorpe}}]{kumar_ring_2014}%
  \BibitemOpen
  \bibfield  {author} {\bibinfo {author} {\bibfnamefont {A.}~\bibnamefont
  {Kumar}}, \bibinfo {author} {\bibfnamefont {D.}~\bibnamefont {Sherrington}},
  \bibinfo {author} {\bibfnamefont {M.}~\bibnamefont {Wilson}},\ and\ \bibinfo
  {author} {\bibfnamefont {M.~F.}\ \bibnamefont {Thorpe}},\ }\href
  {https://doi.org/10.1088/0953-8984/26/39/395401} {\bibfield  {journal}
  {\bibinfo  {journal} {J. Phys.: Condens. Matter}\ }\textbf {\bibinfo {volume}
  {26}},\ \bibinfo {pages} {395401} (\bibinfo {year} {2014})}\BibitemShut
  {NoStop}%
\bibitem [{\citenamefont {Newville}\ \emph {et~al.}(2014)\citenamefont
  {Newville}, \citenamefont {Stensitzki}, \citenamefont {Allen},\ and\
  \citenamefont {Ingargiola}}]{newville_lmfit_2014}%
  \BibitemOpen
  \bibfield  {author} {\bibinfo {author} {\bibfnamefont {M.}~\bibnamefont
  {Newville}}, \bibinfo {author} {\bibfnamefont {T.}~\bibnamefont
  {Stensitzki}}, \bibinfo {author} {\bibfnamefont {D.~B.}\ \bibnamefont
  {Allen}},\ and\ \bibinfo {author} {\bibfnamefont {A.}~\bibnamefont
  {Ingargiola}},\ }\href {https://doi.org/10.5281/zenodo.11813} {\bibinfo
  {title} {{LMFIT}: {Non}-{Linear} {Least}-{Square} {Minimization} and
  {Curve}-{Fitting} for {Python}}} (\bibinfo {year} {2014})\BibitemShut
  {NoStop}%
\bibitem [{\citenamefont {Le~Roux}\ and\ \citenamefont
  {Jund}(2010)}]{le_roux_ring_2010}%
  \BibitemOpen
  \bibfield  {author} {\bibinfo {author} {\bibfnamefont {S.}~\bibnamefont
  {Le~Roux}}\ and\ \bibinfo {author} {\bibfnamefont {P.}~\bibnamefont {Jund}},\
  }\href {https://doi.org/10.1016/j.commatsci.2010.04.023} {\bibfield
  {journal} {\bibinfo  {journal} {Comput. Mat. Sci.}\ }\textbf {\bibinfo
  {volume} {49}},\ \bibinfo {pages} {70} (\bibinfo {year} {2010})}\BibitemShut
  {NoStop}%
\bibitem [{\citenamefont {Aboav}(1970)}]{aboav_arrangement_1970}%
  \BibitemOpen
  \bibfield  {author} {\bibinfo {author} {\bibfnamefont {D.~A.}\ \bibnamefont
  {Aboav}},\ }\href {https://doi.org/10.1016/0026-0800(70)90038} {\bibfield
  {journal} {\bibinfo  {journal} {Metallography}\ }\textbf {\bibinfo {volume}
  {3}},\ \bibinfo {pages} {383} (\bibinfo {year} {1970})}\BibitemShut {NoStop}%
\bibitem [{\citenamefont {Ebrahem}\ \emph {et~al.}(2020)\citenamefont
  {Ebrahem}, \citenamefont {Bamer},\ and\ \citenamefont
  {Markert}}]{ebrahem_vitreous_2020}%
  \BibitemOpen
  \bibfield  {author} {\bibinfo {author} {\bibfnamefont {F.}~\bibnamefont
  {Ebrahem}}, \bibinfo {author} {\bibfnamefont {F.}~\bibnamefont {Bamer}},\
  and\ \bibinfo {author} {\bibfnamefont {B.}~\bibnamefont {Markert}},\ }\href
  {https://doi.org/10.1016/j.msea.2020.139189} {\bibfield  {journal} {\bibinfo
  {journal} {Mat. Sci. Eng.: A}\ }\textbf {\bibinfo {volume} {780}},\ \bibinfo
  {pages} {139189} (\bibinfo {year} {2020})}\BibitemShut {NoStop}%
\bibitem [{\citenamefont {Weaire}\ and\ \citenamefont
  {Rivier}(1984)}]{weaire_soap_1984}%
  \BibitemOpen
  \bibfield  {author} {\bibinfo {author} {\bibfnamefont {D.}~\bibnamefont
  {Weaire}}\ and\ \bibinfo {author} {\bibfnamefont {N.}~\bibnamefont
  {Rivier}},\ }\href {https://doi.org/10.1080/00107518408210979} {\bibfield
  {journal} {\bibinfo  {journal} {Contemp. Phys.}\ }\textbf {\bibinfo {volume}
  {25}},\ \bibinfo {pages} {59} (\bibinfo {year} {1984})}\BibitemShut {NoStop}%
\bibitem [{\citenamefont {Newman}(2018)}]{newman_networks_2018}%
  \BibitemOpen
  \bibfield  {author} {\bibinfo {author} {\bibfnamefont {M.~E.~J.}\
  \bibnamefont {Newman}},\ }\href@noop {} {\emph {\bibinfo {title}
  {Networks}}},\ \bibinfo {edition} {2nd}\ ed.\ (\bibinfo  {publisher} {Oxford
  University Press},\ \bibinfo {year} {2018})\BibitemShut {NoStop}%
\bibitem [{\citenamefont {Saika-Voivod}\ \emph {et~al.}(2001)\citenamefont
  {Saika-Voivod}, \citenamefont {Poole},\ and\ \citenamefont
  {Sciortino}}]{Saika-Voivod_Poole_Sciortino_2001}%
  \BibitemOpen
  \bibfield  {author} {\bibinfo {author} {\bibfnamefont {I.}~\bibnamefont
  {Saika-Voivod}}, \bibinfo {author} {\bibfnamefont {P.~H.}\ \bibnamefont
  {Poole}},\ and\ \bibinfo {author} {\bibfnamefont {F.}~\bibnamefont
  {Sciortino}},\ }\href {https://doi.org/10.1038/35087524} {\bibfield
  {journal} {\bibinfo  {journal} {Nature}\ }\textbf {\bibinfo {volume} {412}},\
  \bibinfo {pages} {514} (\bibinfo {year} {2001})}\BibitemShut {NoStop}%
\bibitem [{\citenamefont {Saika-Voivod}\ \emph {et~al.}(2004)\citenamefont
  {Saika-Voivod}, \citenamefont {Sciortino},\ and\ \citenamefont
  {Poole}}]{Saika-Voivod_Sciortino_Poole_2004}%
  \BibitemOpen
  \bibfield  {author} {\bibinfo {author} {\bibfnamefont {I.}~\bibnamefont
  {Saika-Voivod}}, \bibinfo {author} {\bibfnamefont {F.}~\bibnamefont
  {Sciortino}},\ and\ \bibinfo {author} {\bibfnamefont {P.~H.}\ \bibnamefont
  {Poole}},\ }\href {https://doi.org/10.1103/PhysRevE.69.041503} {\bibfield
  {journal} {\bibinfo  {journal} {Phys. Rev. E}\ }\textbf {\bibinfo {volume}
  {69}},\ \bibinfo {pages} {041503} (\bibinfo {year} {2004})}\BibitemShut
  {NoStop}%
\bibitem [{\citenamefont {Tanaka}(2022)}]{Tanaka_2022}%
  \BibitemOpen
  \bibfield  {author} {\bibinfo {author} {\bibfnamefont {H.}~\bibnamefont
  {Tanaka}},\ }\href {https://doi.org/10.1016/j.nocx.2021.100076} {\bibfield
  {journal} {\bibinfo  {journal} {J. Non-Cryst. Solids: X}\ }\textbf {\bibinfo
  {volume} {13}},\ \bibinfo {pages} {100076} (\bibinfo {year}
  {2022})}\BibitemShut {NoStop}%
\bibitem [{\citenamefont {Heuer}(2008)}]{Heuer_2008}%
  \BibitemOpen
  \bibfield  {author} {\bibinfo {author} {\bibfnamefont {A.}~\bibnamefont
  {Heuer}},\ }\href {https://doi.org/10.1088/0953-8984/20/37/373101} {\bibfield
   {journal} {\bibinfo  {journal} {J. Phys.: Condens. Matt.}\ }\textbf
  {\bibinfo {volume} {20}},\ \bibinfo {pages} {373101} (\bibinfo {year}
  {2008})}\BibitemShut {NoStop}%
\bibitem [{\citenamefont {Stillinger}\ and\ \citenamefont
  {Weber}(1982)}]{Stillinger_Weber_1982}%
  \BibitemOpen
  \bibfield  {author} {\bibinfo {author} {\bibfnamefont {F.~H.}\ \bibnamefont
  {Stillinger}}\ and\ \bibinfo {author} {\bibfnamefont {T.~A.}\ \bibnamefont
  {Weber}},\ }\href {https://doi.org/212} {\bibfield  {journal} {\bibinfo
  {journal} {Phys. Rev. A}\ }\textbf {\bibinfo {volume} {25}},\ \bibinfo
  {pages} {978} (\bibinfo {year} {1982})}\BibitemShut {NoStop}%
\bibitem [{\citenamefont {Sastry}\ \emph {et~al.}(1998)\citenamefont {Sastry},
  \citenamefont {Debenedetti},\ and\ \citenamefont
  {Stillinger}}]{Sastry_Debenedetti_Stillinger_1998}%
  \BibitemOpen
  \bibfield  {author} {\bibinfo {author} {\bibfnamefont {S.}~\bibnamefont
  {Sastry}}, \bibinfo {author} {\bibfnamefont {P.~G.}\ \bibnamefont
  {Debenedetti}},\ and\ \bibinfo {author} {\bibfnamefont {F.~H.}\ \bibnamefont
  {Stillinger}},\ }\href {https://doi.org/10.1038/31189} {\bibfield  {journal}
  {\bibinfo  {journal} {Nature}\ }\textbf {\bibinfo {volume} {393}},\ \bibinfo
  {pages} {554} (\bibinfo {year} {1998})}\BibitemShut {NoStop}%
\bibitem [{\citenamefont {Heuer}\ and\ \citenamefont
  {Büchner}(2000)}]{Heuer_Buchner_2000}%
  \BibitemOpen
  \bibfield  {author} {\bibinfo {author} {\bibfnamefont {A.}~\bibnamefont
  {Heuer}}\ and\ \bibinfo {author} {\bibfnamefont {S.}~\bibnamefont
  {Büchner}},\ }\href {https://doi.org/10.1088/0953-8984/12/29/325} {\bibfield
   {journal} {\bibinfo  {journal} {J. Phys.: Condens. Matter}\ }\textbf
  {\bibinfo {volume} {12}},\ \bibinfo {pages} {6535} (\bibinfo {year}
  {2000})}\BibitemShut {NoStop}%
\bibitem [{Note1()}]{Note1}%
  \BibitemOpen
  \bibinfo {note} {Note that, according to the Mermin-Wagner theorem~\cite
  {Mermin_1968}, the crystalline phase can only display quasi-long range
  order}\BibitemShut {NoStop}%
\bibitem [{\citenamefont {Saksaengwijit}\ and\ \citenamefont
  {Heuer}(2007)}]{Saksaengwijit_Heuer_2007}%
  \BibitemOpen
  \bibfield  {author} {\bibinfo {author} {\bibfnamefont {A.}~\bibnamefont
  {Saksaengwijit}}\ and\ \bibinfo {author} {\bibfnamefont {A.}~\bibnamefont
  {Heuer}},\ }\href {https://doi.org/10.1088/0953-8984/19/20/205143} {\bibfield
   {journal} {\bibinfo  {journal} {J. Phys.: Condens. Matt.}\ }\textbf
  {\bibinfo {volume} {19}},\ \bibinfo {pages} {205143} (\bibinfo {year}
  {2007})}\BibitemShut {NoStop}%
\bibitem [{\citenamefont {Li}\ \emph {et~al.}(2019)\citenamefont {Li},
  \citenamefont {Mishra}, \citenamefont {Sun}, \citenamefont {Zhao},
  \citenamefont {Mason}, \citenamefont {Ganapathy},\ and\ \citenamefont
  {Pica~Ciamarra}}]{li_long-wavelength_2019}%
  \BibitemOpen
  \bibfield  {author} {\bibinfo {author} {\bibfnamefont {Y.-W.}\ \bibnamefont
  {Li}}, \bibinfo {author} {\bibfnamefont {C.~K.}\ \bibnamefont {Mishra}},
  \bibinfo {author} {\bibfnamefont {Z.-Y.}\ \bibnamefont {Sun}}, \bibinfo
  {author} {\bibfnamefont {K.}~\bibnamefont {Zhao}}, \bibinfo {author}
  {\bibfnamefont {T.~G.}\ \bibnamefont {Mason}}, \bibinfo {author}
  {\bibfnamefont {R.}~\bibnamefont {Ganapathy}},\ and\ \bibinfo {author}
  {\bibfnamefont {M.}~\bibnamefont {Pica~Ciamarra}},\ }\href
  {https://doi.org/10.1073/pnas.1909319116} {\bibfield  {journal} {\bibinfo
  {journal} {Proc. Natl. Acad. Sci.}\ }\textbf {\bibinfo {volume} {116}},\
  \bibinfo {pages} {22977} (\bibinfo {year} {2019})}\BibitemShut {NoStop}%
\bibitem [{\citenamefont {Scalliet}\ \emph {et~al.}(2022)\citenamefont
  {Scalliet}, \citenamefont {Guiselin},\ and\ \citenamefont
  {Berthier}}]{Scalliet_Guiselin_Berthier_2022}%
  \BibitemOpen
  \bibfield  {author} {\bibinfo {author} {\bibfnamefont {C.}~\bibnamefont
  {Scalliet}}, \bibinfo {author} {\bibfnamefont {B.}~\bibnamefont {Guiselin}},\
  and\ \bibinfo {author} {\bibfnamefont {L.}~\bibnamefont {Berthier}},\ }\href
  {https://doi.org/10.1103/PhysRevX.12.041028} {\bibfield  {journal} {\bibinfo
  {journal} {Phys. Rev. X}\ }\textbf {\bibinfo {volume} {12}},\ \bibinfo
  {pages} {041028} (\bibinfo {year} {2022})}\BibitemShut {NoStop}%
\bibitem [{\citenamefont {Shi}\ and\ \citenamefont
  {Tanaka}(2018)}]{Shi_Tanaka_2018}%
  \BibitemOpen
  \bibfield  {author} {\bibinfo {author} {\bibfnamefont {R.}~\bibnamefont
  {Shi}}\ and\ \bibinfo {author} {\bibfnamefont {H.}~\bibnamefont {Tanaka}},\
  }\href {https://doi.org/10.1073/pnas.1717233115} {\bibfield  {journal}
  {\bibinfo  {journal} {Proc. Natl. Acad. Sci.}\ }\textbf {\bibinfo {volume}
  {115}},\ \bibinfo {pages} {1980} (\bibinfo {year} {2018})}\BibitemShut
  {NoStop}%
\bibitem [{\citenamefont {Wright}(2020)}]{Wright_2020}%
  \BibitemOpen
  \bibfield  {author} {\bibinfo {author} {\bibfnamefont {A.~C.}\ \bibnamefont
  {Wright}},\ }\href
  {https://www.ingentaconnect.com/content/10.13036/17533562.61.2.02} {\bibfield
   {journal} {\bibinfo  {journal} {Phys. Chem. Glasses: Eur. J. Glass Sci.
  Technol. B}\ }\textbf {\bibinfo {volume} {61}},\ \bibinfo {pages} {57}
  (\bibinfo {year} {2020})}\BibitemShut {NoStop}%
\bibitem [{\citenamefont {Pedersen}\ \emph {et~al.}(2021)\citenamefont
  {Pedersen}, \citenamefont {Douglass},\ and\ \citenamefont
  {Harrowell}}]{Pedersen_Douglass_Harrowell_2021}%
  \BibitemOpen
  \bibfield  {author} {\bibinfo {author} {\bibfnamefont {U.~R.}\ \bibnamefont
  {Pedersen}}, \bibinfo {author} {\bibfnamefont {I.}~\bibnamefont {Douglass}},\
  and\ \bibinfo {author} {\bibfnamefont {P.}~\bibnamefont {Harrowell}},\ }\href
  {https://doi.org/10.1063/5.0033206} {\bibfield  {journal} {\bibinfo
  {journal} {J. Chem. Phys.}\ }\textbf {\bibinfo {volume} {154}},\ \bibinfo
  {pages} {054503} (\bibinfo {year} {2021})}\BibitemShut {NoStop}%
\bibitem [{\citenamefont {Hu}\ \emph {et~al.}(2017)\citenamefont {Hu},
  \citenamefont {Tanaka},\ and\ \citenamefont {Wang}}]{Hu_Tanaka_Wang_2017}%
  \BibitemOpen
  \bibfield  {author} {\bibinfo {author} {\bibfnamefont {Y.-C.}\ \bibnamefont
  {Hu}}, \bibinfo {author} {\bibfnamefont {H.}~\bibnamefont {Tanaka}},\ and\
  \bibinfo {author} {\bibfnamefont {W.-H.}\ \bibnamefont {Wang}},\ }\href
  {https://doi.org/10.1103/PhysRevE.96.022613} {\bibfield  {journal} {\bibinfo
  {journal} {Phys. Rev. E}\ }\textbf {\bibinfo {volume} {96}},\ \bibinfo
  {pages} {022613} (\bibinfo {year} {2017})}\BibitemShut {NoStop}%
\bibitem [{\citenamefont {Massa}\ \emph {et~al.}(2019)\citenamefont {Massa},
  \citenamefont {Leporini},\ and\ \citenamefont
  {Puosi}}]{Massa_Leporini_Puosi_2019}%
  \BibitemOpen
  \bibfield  {author} {\bibinfo {author} {\bibfnamefont {C.~A.}\ \bibnamefont
  {Massa}}, \bibinfo {author} {\bibfnamefont {D.}~\bibnamefont {Leporini}},\
  and\ \bibinfo {author} {\bibfnamefont {F.}~\bibnamefont {Puosi}},\ }\href
  {https://doi.org/10.1088/1361-648X/ab539c} {\bibfield  {journal} {\bibinfo
  {journal} {J. Phys.: Condens. Matt.}\ }\textbf {\bibinfo {volume} {32}},\
  \bibinfo {pages} {085701} (\bibinfo {year} {2019})}\BibitemShut {NoStop}%
\bibitem [{\citenamefont {Niu}\ \emph {et~al.}(2018)\citenamefont {Niu},
  \citenamefont {Piaggi}, \citenamefont {Invernizzi},\ and\ \citenamefont
  {Parrinello}}]{Niu_Piaggi_Invernizzi_Parrinello_2018}%
  \BibitemOpen
  \bibfield  {author} {\bibinfo {author} {\bibfnamefont {H.}~\bibnamefont
  {Niu}}, \bibinfo {author} {\bibfnamefont {P.~M.}\ \bibnamefont {Piaggi}},
  \bibinfo {author} {\bibfnamefont {M.}~\bibnamefont {Invernizzi}},\ and\
  \bibinfo {author} {\bibfnamefont {M.}~\bibnamefont {Parrinello}},\ }\href
  {https://doi.org/10.1073/pnas.1803919115} {\bibfield  {journal} {\bibinfo
  {journal} {Proc. Natl. Acad. Sci.}\ }\textbf {\bibinfo {volume} {115}},\
  \bibinfo {pages} {5348} (\bibinfo {year} {2018})}\BibitemShut {NoStop}%
\bibitem [{\citenamefont {Tong}\ and\ \citenamefont
  {Tanaka}(2023)}]{Tong_Tanaka_2023}%
  \BibitemOpen
  \bibfield  {author} {\bibinfo {author} {\bibfnamefont {H.}~\bibnamefont
  {Tong}}\ and\ \bibinfo {author} {\bibfnamefont {H.}~\bibnamefont {Tanaka}},\
  }\href {https://doi.org/10.1038/s41467-023} {\bibfield  {journal} {\bibinfo
  {journal} {Nat. Commun.}\ }\textbf {\bibinfo {volume} {14}},\ \bibinfo
  {pages} {4614} (\bibinfo {year} {2023})}\BibitemShut {NoStop}%
\bibitem [{\citenamefont {Coslovich}\ \emph {et~al.}(2018)\citenamefont
  {Coslovich}, \citenamefont {Ozawa},\ and\ \citenamefont
  {Berthier}}]{Coslovich_Ozawa_Berthier_2018}%
  \BibitemOpen
  \bibfield  {author} {\bibinfo {author} {\bibfnamefont {D.}~\bibnamefont
  {Coslovich}}, \bibinfo {author} {\bibfnamefont {M.}~\bibnamefont {Ozawa}},\
  and\ \bibinfo {author} {\bibfnamefont {L.}~\bibnamefont {Berthier}},\ }\href
  {https://doi.org/10.1088/1361-648X/aab0c9} {\bibfield  {journal} {\bibinfo
  {journal} {J. Phys.: Condens. Matt.}\ }\textbf {\bibinfo {volume} {30}},\
  \bibinfo {pages} {144004} (\bibinfo {year} {2018})}\BibitemShut {NoStop}%
\bibitem [{\citenamefont {Xu}\ \emph {et~al.}(2005)\citenamefont {Xu},
  \citenamefont {Kumar}, \citenamefont {Buldyrev}, \citenamefont {Chen},
  \citenamefont {Poole}, \citenamefont {Sciortino},\ and\ \citenamefont
  {Stanley}}]{Xu_Kumar_Buldyrev_Chen_Poole_Sciortino_Stanley_2005}%
  \BibitemOpen
  \bibfield  {author} {\bibinfo {author} {\bibfnamefont {L.}~\bibnamefont
  {Xu}}, \bibinfo {author} {\bibfnamefont {P.}~\bibnamefont {Kumar}}, \bibinfo
  {author} {\bibfnamefont {S.~V.}\ \bibnamefont {Buldyrev}}, \bibinfo {author}
  {\bibfnamefont {S.-H.}\ \bibnamefont {Chen}}, \bibinfo {author}
  {\bibfnamefont {P.~H.}\ \bibnamefont {Poole}}, \bibinfo {author}
  {\bibfnamefont {F.}~\bibnamefont {Sciortino}},\ and\ \bibinfo {author}
  {\bibfnamefont {H.~E.}\ \bibnamefont {Stanley}},\ }\href
  {https://doi.org/10.1073/pnas.0507870102} {\bibfield  {journal} {\bibinfo
  {journal} {Proc. Natl. Acad. Sci.}\ }\textbf {\bibinfo {volume} {102}},\
  \bibinfo {pages} {16558} (\bibinfo {year} {2005})}\BibitemShut {NoStop}%
\bibitem [{\citenamefont {Coslovich}\ \emph {et~al.}(2019)\citenamefont
  {Coslovich}, \citenamefont {Ninarello},\ and\ \citenamefont
  {Berthier}}]{Coslovich_Ninarello_Berthier_2019}%
  \BibitemOpen
  \bibfield  {author} {\bibinfo {author} {\bibfnamefont {D.}~\bibnamefont
  {Coslovich}}, \bibinfo {author} {\bibfnamefont {A.}~\bibnamefont
  {Ninarello}},\ and\ \bibinfo {author} {\bibfnamefont {L.}~\bibnamefont
  {Berthier}},\ }\href {https://doi.org/10.21468/SciPostPhys.7.6.077}
  {\bibfield  {journal} {\bibinfo  {journal} {SciPost Physics}\ }\textbf
  {\bibinfo {volume} {7}},\ \bibinfo {pages} {077} (\bibinfo {year}
  {2019})}\BibitemShut {NoStop}%
\bibitem [{\citenamefont {Luo}\ \emph {et~al.}(2022)\citenamefont {Luo},
  \citenamefont {Robinson}, \citenamefont {Pihlajamaa}, \citenamefont {Debets},
  \citenamefont {Royall},\ and\ \citenamefont
  {Janssen}}]{Luo_Robinson_Pihlajamaa_Debets_Royall_Janssen_2022}%
  \BibitemOpen
  \bibfield  {author} {\bibinfo {author} {\bibfnamefont {C.}~\bibnamefont
  {Luo}}, \bibinfo {author} {\bibfnamefont {J.~F.}\ \bibnamefont {Robinson}},
  \bibinfo {author} {\bibfnamefont {I.}~\bibnamefont {Pihlajamaa}}, \bibinfo
  {author} {\bibfnamefont {V.~E.}\ \bibnamefont {Debets}}, \bibinfo {author}
  {\bibfnamefont {C.~P.}\ \bibnamefont {Royall}},\ and\ \bibinfo {author}
  {\bibfnamefont {L.~M.~C.}\ \bibnamefont {Janssen}},\ }\href
  {https://doi.org/10.1103/PhysRevLett.129.145501} {\bibfield  {journal}
  {\bibinfo  {journal} {Phys. Rev. L}\ }\textbf {\bibinfo {volume} {129}},\
  \bibinfo {pages} {145501} (\bibinfo {year} {2022})}\BibitemShut {NoStop}%
\bibitem [{\citenamefont {Shi}\ \emph {et~al.}(2018)\citenamefont {Shi},
  \citenamefont {Russo},\ and\ \citenamefont {Tanaka}}]{Shi_Russo_Tanaka_2018}%
  \BibitemOpen
  \bibfield  {author} {\bibinfo {author} {\bibfnamefont {R.}~\bibnamefont
  {Shi}}, \bibinfo {author} {\bibfnamefont {J.}~\bibnamefont {Russo}},\ and\
  \bibinfo {author} {\bibfnamefont {H.}~\bibnamefont {Tanaka}},\ }\href
  {https://doi.org/10.1073/pnas.1807821115} {\bibfield  {journal} {\bibinfo
  {journal} {Proc. Natl. Acad. Sci.}\ }\textbf {\bibinfo {volume} {115}},\
  \bibinfo {pages} {9444} (\bibinfo {year} {2018})}\BibitemShut {NoStop}%
\bibitem [{\citenamefont {Ozawa}\ \emph {et~al.}(2019)\citenamefont {Ozawa},
  \citenamefont {Scalliet}, \citenamefont {Ninarello},\ and\ \citenamefont
  {Berthier}}]{Ozawa_Scalliet_Ninarello_Berthier_2019}%
  \BibitemOpen
  \bibfield  {author} {\bibinfo {author} {\bibfnamefont {M.}~\bibnamefont
  {Ozawa}}, \bibinfo {author} {\bibfnamefont {C.}~\bibnamefont {Scalliet}},
  \bibinfo {author} {\bibfnamefont {A.}~\bibnamefont {Ninarello}},\ and\
  \bibinfo {author} {\bibfnamefont {L.}~\bibnamefont {Berthier}},\ }\href
  {https://doi.org/10.1063/1.5113477} {\bibfield  {journal} {\bibinfo
  {journal} {J. Chem. Phys.}\ }\textbf {\bibinfo {volume} {151}},\ \bibinfo
  {pages} {084504} (\bibinfo {year} {2019})}\BibitemShut {NoStop}%
\bibitem [{\citenamefont {Widmer-Cooper}\ \emph {et~al.}(2004)\citenamefont
  {Widmer-Cooper}, \citenamefont {Harrowell},\ and\ \citenamefont
  {Fynewever}}]{Widmer-Cooper_Harrowell_Fynewever_2004}%
  \BibitemOpen
  \bibfield  {author} {\bibinfo {author} {\bibfnamefont {A.}~\bibnamefont
  {Widmer-Cooper}}, \bibinfo {author} {\bibfnamefont {P.}~\bibnamefont
  {Harrowell}},\ and\ \bibinfo {author} {\bibfnamefont {H.}~\bibnamefont
  {Fynewever}},\ }\href {https://doi.org/10.1103/PhysRevLett.93.135701}
  {\bibfield  {journal} {\bibinfo  {journal} {Phys. Rev. Lett.}\ }\textbf
  {\bibinfo {volume} {93}},\ \bibinfo {pages} {135701} (\bibinfo {year}
  {2004})}\BibitemShut {NoStop}%
\bibitem [{\citenamefont {Mermin}(1968)}]{Mermin_1968}%
  \BibitemOpen
  \bibfield  {author} {\bibinfo {author} {\bibfnamefont {N.~D.}\ \bibnamefont
  {Mermin}},\ }\href {https://doi.org/10.1103/PhysRev.176.250} {\bibfield
  {journal} {\bibinfo  {journal} {Phys. Rev.}\ }\textbf {\bibinfo {volume}
  {176}},\ \bibinfo {pages} {250} (\bibinfo {year} {1968})}\BibitemShut
  {NoStop}%
\bibitem [{\citenamefont {Sadjadi}\ \emph {et~al.}(2017)\citenamefont
  {Sadjadi}, \citenamefont {Bhattarai}, \citenamefont {Drabold}, \citenamefont
  {Thorpe},\ and\ \citenamefont
  {Wilson}}]{Sadjadi_Bhattarai_Drabold_Thorpe_Wilson_2017}%
  \BibitemOpen
  \bibfield  {author} {\bibinfo {author} {\bibfnamefont {M.}~\bibnamefont
  {Sadjadi}}, \bibinfo {author} {\bibfnamefont {B.}~\bibnamefont {Bhattarai}},
  \bibinfo {author} {\bibfnamefont {D.~A.}\ \bibnamefont {Drabold}}, \bibinfo
  {author} {\bibfnamefont {M.~F.}\ \bibnamefont {Thorpe}},\ and\ \bibinfo
  {author} {\bibfnamefont {M.}~\bibnamefont {Wilson}},\ }\href
  {https://doi.org/10.1103/PhysRevB.96.201405} {\bibfield  {journal} {\bibinfo
  {journal} {Phys. Rev. B}\ }\textbf {\bibinfo {volume} {96}},\ \bibinfo
  {pages} {201405} (\bibinfo {year} {2017})}\BibitemShut {NoStop}%
\bibitem [{\citenamefont {Guttman}(1990)}]{guttman_ring_1990}%
  \BibitemOpen
  \bibfield  {author} {\bibinfo {author} {\bibfnamefont {L.}~\bibnamefont
  {Guttman}},\ }\href {https://doi.org/10.1016/0022-3093(90)90686-G} {\bibfield
   {journal} {\bibinfo  {journal} {J. Non-Cryst. Solids}\ }\textbf {\bibinfo
  {volume} {116}},\ \bibinfo {pages} {145} (\bibinfo {year}
  {1990})}\BibitemShut {NoStop}%
\bibitem [{\citenamefont {King}(1967)}]{king_ring_1967}%
  \BibitemOpen
  \bibfield  {author} {\bibinfo {author} {\bibfnamefont {S.~V.}\ \bibnamefont
  {King}},\ }\href {https://doi.org/10.1038/2131112a0} {\bibfield  {journal}
  {\bibinfo  {journal} {Nature}\ }\textbf {\bibinfo {volume} {213}},\ \bibinfo
  {pages} {1112} (\bibinfo {year} {1967})}\BibitemShut {NoStop}%
\bibitem [{\citenamefont {Franzblau}(1991)}]{franzblau_computation_1991}%
  \BibitemOpen
  \bibfield  {author} {\bibinfo {author} {\bibfnamefont {D.~S.}\ \bibnamefont
  {Franzblau}},\ }\href {https://doi.org/10.1103/PhysRevB.44.4925} {\bibfield
  {journal} {\bibinfo  {journal} {Phys. Rev. B}\ }\textbf {\bibinfo {volume}
  {44}},\ \bibinfo {pages} {4925} (\bibinfo {year} {1991})}\BibitemShut
  {NoStop}%
\bibitem [{\citenamefont {Goetzke}\ and\ \citenamefont
  {Klein}(1991)}]{goetzke_properties_1991}%
  \BibitemOpen
  \bibfield  {author} {\bibinfo {author} {\bibfnamefont {K.}~\bibnamefont
  {Goetzke}}\ and\ \bibinfo {author} {\bibfnamefont {H.~J.}\ \bibnamefont
  {Klein}},\ }\href {https://doi.org/10.1016/0022-3093(91)90145-V} {\bibfield
  {journal} {\bibinfo  {journal} {J. Non-Cryst. Solids}\ }\textbf {\bibinfo
  {volume} {127}},\ \bibinfo {pages} {215} (\bibinfo {year}
  {1991})}\BibitemShut {NoStop}%
\bibitem [{\citenamefont {Wooten}(2002)}]{wooten_structure_2002}%
  \BibitemOpen
  \bibfield  {author} {\bibinfo {author} {\bibfnamefont {F.}~\bibnamefont
  {Wooten}},\ }\href {https://doi.org/10.1107/S0108767302006669} {\bibfield
  {journal} {\bibinfo  {journal} {Acta Cryst A}\ }\textbf {\bibinfo {volume}
  {58}},\ \bibinfo {pages} {346} (\bibinfo {year} {2002})}\BibitemShut
  {NoStop}%
\bibitem [{\citenamefont {Yuan}\ and\ \citenamefont
  {Cormack}(2002)}]{yuan_efficient_2002}%
  \BibitemOpen
  \bibfield  {author} {\bibinfo {author} {\bibfnamefont {X.}~\bibnamefont
  {Yuan}}\ and\ \bibinfo {author} {\bibfnamefont {A.~N.}\ \bibnamefont
  {Cormack}},\ }\href {https://doi.org/10.1016/S0927-0256(01)00256} {\bibfield
  {journal} {\bibinfo  {journal} {Comp. Mat. Sci.}\ }\textbf {\bibinfo {volume}
  {24}},\ \bibinfo {pages} {343} (\bibinfo {year} {2002})}\BibitemShut
  {NoStop}%
\bibitem [{\citenamefont {Tirelli}\ and\ \citenamefont
  {Nakano}(2023)}]{tirelli_topological_2023}%
  \BibitemOpen
  \bibfield  {author} {\bibinfo {author} {\bibfnamefont {A.}~\bibnamefont
  {Tirelli}}\ and\ \bibinfo {author} {\bibfnamefont {K.}~\bibnamefont
  {Nakano}},\ }\href {https://doi.org/10.1021/acs.jpcb.2c09009} {\bibfield
  {journal} {\bibinfo  {journal} {J. Phys. Chem. B}\ }\textbf {\bibinfo
  {volume} {127}},\ \bibinfo {pages} {3302} (\bibinfo {year}
  {2023})}\BibitemShut {NoStop}%
\bibitem [{\citenamefont {Trudeau}(1993)}]{trudeau_introduction_1993}%
  \BibitemOpen
  \bibfield  {author} {\bibinfo {author} {\bibfnamefont {R.~J.}\ \bibnamefont
  {Trudeau}},\ }\href@noop {} {\emph {\bibinfo {title} {Introduction to graph
  theory}}},\ \bibinfo {edition} {2nd}\ ed.\ (\bibinfo  {publisher} {Dover
  Publications},\ \bibinfo {year} {1993})\BibitemShut {NoStop}%
\bibitem [{\citenamefont {Dirindin}(2024)}]{l2dr}%
  \BibitemOpen
  \bibfield  {author} {\bibinfo {author} {\bibfnamefont {M.}~\bibnamefont
  {Dirindin}},\ }\href {https://doi.org/10.5281/zenodo.13913245} {\bibinfo
  {title} {{l2dr}}} (\bibinfo {year} {2024})\BibitemShut {NoStop}%
\end{thebibliography}

%

\end{document}